\documentclass[lettersize,journal]{IEEEtran}
\usepackage{amsmath,amssymb,amsfonts}
\usepackage{algorithmic}
\usepackage{array}
\usepackage{textcomp}
\usepackage{stfloats}
\usepackage{url}
\usepackage{subfigure}
\usepackage{verbatim}
\usepackage{graphicx}
\usepackage{cite}

\usepackage{wrapfig}
\usepackage{subfloat}
\usepackage{float}
\usepackage{ulem}
\usepackage{epstopdf}

\def\BibTeX{{\rm B\kern-.05em{\sc i\kern-.025em b}\kern-.08em
    T\kern-.1667em\lower.7ex\hbox{E}\kern-.125emX}}
\begin{document}
\title{Performance Optimization and Parameters Estimation for MIMO-OFDM Dual-functional Communication-radar Systems}
\author{Chen Zhong, Chunrong Gu, Lan Tang, \IEEEmembership{Member, IEEE}, Yechao Bai and Mengting Lou
\thanks{This work was supported by the National Natural Science Foundation of China under grants 62072229, U1936201, 62071220, 61976113 and joint project of China Mobile Research Institute \& X-NET.}
\thanks{Chen Zhong, Chunrong Gu, Lan Tang, and Yechao Bai are with the School of Electronic Science and Engineering, Nanjing University, Nanjing210093, P.R.China.(email:tanglan@nju.edu.cn).}
\thanks{Mengting Lou is with Future Research Lab, China Mobile Research Institute, Beijing 100053, China.}}

\markboth{}%
{Performance Optimization and Parameters Estimation for MIMO-OFDM Dual-functional Communication-radar Systems}

\maketitle

\begin{abstract}
In dual-functional communication-radar systems, common radio frequency (RF) signals are used for both communication and detection. For better compatibility with existing communication systems, we adopt multiple-input multiple-output (MIMO) orthogonal frequency division multiplexing (OFDM) signals as integrated signals and investigate the estimation performance of MIMO-OFDM signals. We first analyze the Cramer-Rao lower bound (CRLB) of parameters estimation. Then, transmit powers over different subcarriers are optimized to achieve the best tradeoff between transmission rate and estimation performance. Finally, we propose a more accurate estimation method which utilizes canonical polyadic decomposition (CPD) of three-order tensor to obtain the parameter matrices. Due to the characteristic of the column structure of the parameter matrices, we just need to use DFT / IDFT to recover the parameters of multiple targets. The simulation results show that the estimation method based on tensor can achieve performance close to CRLB and the estimation performance can be improved by optimizing the transmit powers.
\end{abstract}


\section{Introduction}
\label{1}
\IEEEPARstart{R}{adar}  sensing and wireless communication are two important applications of radio frequency (RF) signals. In the past, they were designed and developed independently over the different frequency bands. Recent years, the demand of ubiquitous perception and overcrowding of spectrum in wireless communication network promote the development of joint communication-radar technology\cite{2016Spectrum}. Meanwhile, the similarity in operating frequency band and signal design makes the communication-radar integration possible.

Integrated communication-radar systems are to realize the communication and sensing on the common hardware platform, which can detect and track targets on the basis of ensuring communication performance. There has been some research work on design of integrated communication-radar systems. These work can be mainly divided into two categories. One is based on the multiplexing technology including the time division multiplexing (TDM) which implements sensing and communication at the different time \cite{2011Multifunctional} and the space division multiplexing (SDM) which utilizes the different subarrays combination for beam division to achieve the different system functions \cite{2017Simultaneous}. The other is based on the integrated waveform design , in which communication bits are modulated on the conventional radar waveform \cite{ 2015Intrapulse, 2016Dual, 2021Performance} or the communication waveforms are modified to satisfy the sensing requirement\cite{2015OFDM, 2017Adaptive, 2018Low}.

In wireless communication systems, OFDM is a key technology because of its resistance to multipath fading, high bandwidth utilization and easy implementation \cite{2006Orthogonal}. To the authors' knowledge, the application of OFDM in the radar system was first proposed in \cite{2000Multifrequency}. Subsequently, many researchers have made active explorations on the OFDM technology for radar system. In terms of the ambiguity function of OFDM, OFDM radar has better range-Doppler ambiguity than traditional linear frequency modulation (LFM) radar \cite{2006Doppler}. In order to improve ranging performance, a direct processing method utilizing OFDM modulation symbols was proposed in \cite{2009An}. In passive radar using OFDM signals, high computational complexity MUSIC-based two-dimensional search algorithm and compressed sensing algorithm were applied in the receiver to obtain better speed-range resolution and clutter removal \cite{2010Signal}. In \cite{2020Super}, an auto-paired super-resolution range and velocity estimation method was proposed by exploiting the OFDM waveform. What's more, \cite{2011Waveform} investigated an intelligent transportation networks based on OFDM signals that can communicate and sense simultaneously, and verified the feasibility of using MIMO technology to estimate the angle of arrival at the receiver. Compared with the traditional phased array radar, MIMO radar has great potential in improving resolution and anti-interference. Therefore, some detection and estimation algorithms based on MIMO-OFDM signals were proposed. In \cite{2018Comparison}, existing direction of arrival (DoA) estimation methods of MIMO-OFDM radar system were summarized and compared. \cite{2016Range} designed an integrated waveform by combining MIMO radar and OFDM communication signal, and proposed the corresponding range-angle estimation method.

So far, most of the integrated designs we have mentioned just utilized communication signals to realize the sensing function of radar. To achieve the best tradeoff performance, the transmission scheme needs to be optimized by taking into account the performance of communication and sensing. By optimizing the ratio of the number of data symbols to the total number of symbols in a standard IEEE 802.11ad frame, the sum of the mean square error (MSE) bounds for velocity estimation, range estimation and communication data estimation was minimized in \cite{2017Performance}. In \cite{2018Power}, a scenario where communication base stations (BSs) and radar receiver share spectrum was considered. The transmit power of OFDM radar was minimized under the constraints of the specified mutual information requirement and the predefined channel capacity requirement. \cite{2020A} investigated a massive MIMO-OFDM vehicle system and utilized different antenna sets to realize the communication function and sensing function respectively. When the communication and radar coexist in the shared spectrum at the same time, the sensing signal may cause interference to communication users. A feasible method to  eliminate the interference is to force the radar signals to fall into the nullspace of the channel between the radar antennas and communication users \cite{2018MU}. Meanwhile, \cite{2018MU} also proposed the other architecture which used one waveform to realize both radar and communication functions.

In this paper, we consider an integrated MIMO-OFDM communication-radar system. In this system, the transmitter transmits beamformed OFDM signals to multiple communication receivers and meanwhile, the radar receiver estimates the locations and speeds of communication receivers utilizing the received echo signals. We first analyze the CRLB of estimated parameters in a multi-target scenario when MIMO-OFDM information symbols are employed for detection. Considering the power allocation over different subcarriers will affect both communication and detection performance, we optimize the power allocation to achieve the best tradeoff between communication and radar performance. In this model, the received signal after matched filter can be formulated as a third-order tensor, which can be recognized as canonical polyadic decomposition (CPD) model, also known as parallel factor analysis (PARAFAC) model. The Alternating Least Squares (ALS) algorithm can be used to estimate the factor matrices for decoupling the DoA, Doppler velocity and range items. According to its structural characteristics, DFT / IDFT is taken to recover the targets' parameters. The simulation results show that compared with the method mentioned in \cite{2010Signal}, the accuracy of the estimation is better.

This paper uses lower-case, bold lower-case and bold upper-case for the scalars, vectors and matrices respectively. $ \left( \cdot \right) ^{*} $, $ \left( \cdot \right) ^{T} $, $ \left( \cdot \right) ^{H} $, $ \left( \cdot \right) ^{\dag} $denote complex conjugate, transpose, conjugate transpose and matrix pseudo inverse respectively. The third-order tensor is represented by $ \mathcal{Y} $ .The notations $ \otimes $, $ \circ $, $ \oplus $ and $ \odot $ are the Kronecker, outer vector, Khatri-Rao products and Hadamard products respectively. $ \Vert \cdot \Vert_{F} $ denotes the Forbenius norm. Finally, $ D_{n} \left(\mathbf A \right) $ returns a diagonal matrix with elements of $ n $-th row of $ \mathbf A $ as the diagonal elements and $ vec(\cdot) $ builds a vector from a matrix by stacking its columns.

\begin{figure}
	\centerline{\includegraphics[scale=0.028]{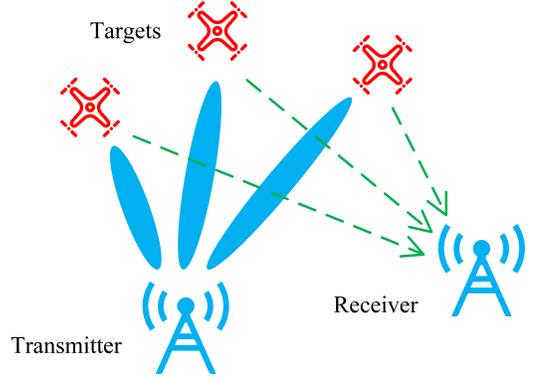}}
	\caption{A bistatic dual-function communication-radar system.}
	\label{fig1}
	\vspace{-10pt}
\end{figure}

\section{System model}\label{2}
As shown in Fig.\ref{fig1}, we consider a bistatic dual-function communication-radar system which can simultaneously communicate with $K$ targets and estimate the parameters of targets utilizing MIMO-OFDM signals. We assume that the transmitter and the receiver are equipped with $ M_{T} $ and $ M_{R} $ antenna elements respectively, and each target only has single antenna. To transmit information bits to $K$ targets, different subcarrier sets are allocated to different targets. We assume that the subcarrier allocation is predetermined. The receiver estimates the target parameters with the received echo signals. The transmitted baseband OFDM signals are divided into $ Q $ blocks. There are $ N $ orthogonal subcarriers in each block and the frequency interval of adjacent subcarriers is $ \Delta f = 1 / T $. The duration of each block is $ T_{s} = T + T_{cp} $ where $ T_{cp} $ is the length of cyclic prefix. The transmitted baseband signal in the $ q $-th data block is given by
\vspace{-5pt}
\begin{equation}
	\begin{aligned}
		\mathbf{x}^{q} ( t ) & =  \sum_{ n=1}^{N}  \mathbf{w}_{ n }  s_{q}(n)    e^{j2\pi\left( n-1 \right) \Delta ft}
		\xi \left( t - \left( q-1 \right)  T_{s} \right), \\
		& \left( q-1 \right) T_{s} - T_{cp} \leq t \leq \left( q - 1 \right) T_{s} + T,
	\end{aligned}
\end{equation}
where $ \mathbf{w}_{n} \in \mathbb{C}^{M_{T} \times 1} $ is the transmit beamforming vector of the $ n $-th subcarrier, $ s_{q} \left( n \right) $ is the transmit symbol on the $n$-th subcarrier in the $ q $-th data block and
\begin{equation}
	\xi \left( t \right) = \left\{
	\begin{aligned}
		& 1,\quad t \in \left[ -T_{cp}, T \right], \\
		& 0,\quad otherwise.
	\end{aligned}	
	\right.
\end{equation}
Therefore, the transmitted signals over $ Q $ blocks can be expressed as $ \mathbf{x} \left( t \right) = \sum_{q=1}^{Q} \mathbf{x}^{q}(t),\ -T_{cp} \leq t \leq \left( Q-1 \right) T_{s} + T $. The baseband signal is then upconverted for transmission, i.e., $ \widetilde{\mathbf{x}}( t ) = {\Re} \left\lbrace  \mathbf{x}(t) e^{j2\pi f_{c}t} \right\rbrace $ where $ f_{c} $ denotes the carrier frequency.

Suppose there are $ K $ targets are in the probing range. We denote $\alpha_{k}$, $ A_{k}^{'} $ and $\varphi_k$ as the direction of departure (DoD), the propagation loss and the phase shift due to the delay and the Doppler frequency of the $ k $-th target respectively. The communication signal received by target $k$ on the $n$-th subcarrier in the $q$-th block is
\vspace{-2pt}
\begin{equation}
	\begin{aligned}
		z_{k}^{q}(n)= \mathbf h^{(q)T}_{k,n} \mathbf w_n s_q(n) + v'(n), \quad n\in \mathcal N_k,
	\end{aligned}
\vspace{-2pt}
\end{equation}
where $ \mathbf h^q_{k,n}= A_{k}^{'}\sqrt{M_T} e^{-j\varphi_k} \mathbf{a}^n_{t} \left( \alpha_{k} \right) $, $ \mathbf{a}^n_{t} \left( \alpha_{k} \right) = \frac{1}{\sqrt{M_{T}}} $ $ \big[1,e^{-j\frac{2\pi}{\lambda_n}d \sin \alpha_{k}}, \dots,e^{-j\frac{2\pi}{\lambda_n}\left( M_{T} - 1\right)d \sin \alpha_{k}} \big] ^{T} $ is the transmit steering vector of the $k$-th target on the $n$-th subcarrier, $ \lambda_n = \frac{c}{f_c+(n-1)\Delta f} $, $c$ is the speed of light, $ d $ is the antenna spacing, $ v'(n) $ is a complex Gaussian variable with zero mean and variance $ n'_0/T $, $ n'_0 $ is the noise power spectral density (PSD) at the communication receiver and $\mathcal N_k$ is the subcarrier set allocated to target $k$. Due to $ \Delta f \ll f_c $, we set $\mathbf{a}^n_{t} \left( \alpha_{k} \right) \approx \mathbf{a}_{t} \left( \alpha_{k} \right) = $ $ \frac{1}{\sqrt{M_{T}}} \big[1,e^{-j\frac{2\pi}{\lambda}d \sin \alpha_{k}}, \dots,e^{-j\frac{2\pi}{\lambda}\left( M_{T} - 1\right)d \sin \alpha_{k}} \big] ^{T}$ in which $ \lambda = \frac{c}{f_c} $ \cite{2009MIMO}. The transmission rate between the transmitter and the $k$th target is
\vspace{-5pt}
\begin{equation}
	\begin{aligned}
		R_k = \sum_{n\in \mathcal N_k}\log\Big(1+\frac{|\mathbf h_{k,n}^{(q)T} \mathbf w_n|^2 p_n T}{n'_0}\Big),
	\end{aligned}
\vspace{-5pt}
\end{equation}
where  $ p_{n} = \mathbb{E}\{\left| s_{q}(n) \right|^{2}\} $ is the transmit power on the $ n $-th subcarrier.

On the other hand, after down-coversion, the echo signals received by the radar receiver can be expressed as
\vspace{-5pt}
\begin{equation}
	\begin{aligned}
		\mathbf{\widetilde y} ( t )
		& = \mathcal{K} \sum_{q=1}^{Q} \sum_{k=1}^{K} \widetilde{A}_{k} e^{j2\pi f_{k}t} \mathbf{a}_{r}(\beta_{k}) \mathbf{a}_{t}^{T}(\alpha_{k})
		\mathbf{x}^{q} \left( t - \tau_{k} \right) + \mathbf{\widetilde v}(t)  \\
		& \approx \mathcal{K} \sum_{q=1}^{Q} \sum_{k=1}^{K} \widetilde{A}_{k} e^{j2\pi f_{k}(q-1)T_{s}} \mathbf{a}_{r}(\beta_{k}) \mathbf{a}_{t}^{T}(\alpha_{k}) \mathbf{x}^{q} \left( t - \tau_{k} \right)\\
		& + \mathbf{\widetilde v}(t),
	\end{aligned}
\vspace{-2pt}
\end{equation}
where $ \mathcal{K} = \sqrt{M_TM_R}$ is the array gain factor, $ \widetilde{A}_{k} $ is the attenuation coefficient including two-propagation loss and reflection coefficient, $ f_{k} = v_{k} / \lambda $ is the Doppler frequency, $ \beta_{k} $ is the DoA, $ \tau_{k} = r_{k} / c = (r_k^1 + r_k^2)/c$ is the two-way delay, $ r_k^1 $ is the distance between the transmitter and the target $ k $, $ r_k^2 $ is the distance between the target $ k $ and the receiver, $ \mathbf{a}_{r} (\beta_{k}) =  \frac{1}{\sqrt{M_{R}}} \big[ 1,e^{-j\frac{2\pi}{\lambda}d \sin \beta_{k}}, \dots,e^{-j\frac{2\pi}{\lambda}\left( M_{R} - 1\right)d \sin \beta_{k}} \big] ^{T} $ is the receive steering vector of the $k$th target and $ \mathbf{\widetilde v}(t) $ is the complex white Gaussian noise with PSD $ n_0 $. The approximation in (5) is because the phase rotation within one OFDM block due to the Doppler frequency can be approximated as a constant when $ f_{k} T_{s} \ll 1 $ \cite{2010Signal}.

After matched filtering, the signal in the $ n $-th subcarrier in the $q$th block can be expressed as
\begin{equation}
	\begin{aligned}
		\mathbf{y}^{q}(n)
		& = \int_{\left( q-1 \right) T_{s}} ^{\left( q-1 \right) T_{s}+T} s_{q}^{*}(n) e^{-j2\pi\left( n-1\right) \Delta ft} \mathbf{\widetilde y}(t)dt \\
		& = \mathcal{K} \sum_{k=1}^{K} \widetilde{A}_{k} T e^{j2\pi f_{k}(q-1)T_{s}} e^{-j2\pi(n-1) \Delta f \tau_{k}} \left| s_{q}(n) \right|^{2} \\
		& \quad  \quad \; \; \times \mathbf{a}_{r}(\beta_{k}) \mathbf{a}_{t}^{T}(\alpha_{k})  \mathbf{w}_{n} + \mathbf{v}^{q}(n),
	\end{aligned}
\end{equation}
where $ \mathbf{v}^{q}(n)
= \int_{\left( q-1 \right) T_{s}} ^{\left( q-1 \right) T_{s}+T}
s_{q}^{*}(n) e^{-j2\pi(n-1) \Delta ft} \mathbf{\widetilde v}(t) dt $.
For the sake of simplicity, we define $ A_{k} = \mathcal{K} \widetilde{A}_{k} T$. $ \mathbf{v}^{q}(n) $ is the complex Gaussian vector with zero  mean and convariance matrix $ p_{n}n_0T \mathbf{I} $ where $\mathbf I$ is the $M_R \times M_R$ unit matrix. Then, vectors in (6) are arranged in the following form of matrix
\vspace{0pt}
\begin{equation}
	\begin{aligned}
		\mathbf{Y}^{(q)} & = \left[ \mathbf{y}^{q} \left( 1 \right) , \mathbf{y}^{q} \left( 2 \right), \dots, \mathbf{y}^{q} \left( N \right) \right] \\
		& = \sum_{k=1}^{K} {A}_{k} e^{j2\pi f_{k}(q-1)T_{s}} \mathbf{a}_{r}\left(\beta_{k} \right)
		\mathbf{a}_{t}^{T}\left(\alpha_{k} \right) \\
		& \times \left[ \mathbf{w}_{ 1 } p_{1}, \mathbf{w}_{ 2 }  p_{2} e^{-j2\pi \Delta f \tau_{k}}, \dots, \mathbf{w}_{ N }  p_{N} e^{-j2\pi \left( N-1 \right) \Delta f \tau_{k}} \right] \\
		& + \left[ \mathbf{v}^{q} \left( 1 \right),\mathbf{v}^{q} \left( 2 \right),\dots, \mathbf{v}^{q} \left( N \right)\right] \\
		& = \mathbf{A}_{R} D_{q}\left( \mathbf{C} \right) \mathbf{B}^{T} + \mathbf{V}^{(q)},
	\end{aligned}
\end{equation}
where $ \mathbf{A}_{R} = \big[\mathbf{a}_{r}(\beta_{1}), \mathbf{a}_{r}(\beta_{2}), \dots, \mathbf{a}_{r}(\beta_{K}) \big] $, $ \mathbf{C} = \big[\mathbf{c}(f_{1}), $ $ \mathbf{c}(f_{2}),\, \dots,\, \mathbf{c}(f_{K})\big] $, $ \mathbf{c}(f_{k})  =  \big[ {A}_{k},\,  {A}_{k} e^{j2\pi f_{k}T_{s}}, \, \dots, \, {A}_{k}  $ $ e^{j2\pi f_{k}(Q-1)T_{s}} \big] ^{T} $, $ \mathbf{B} = \big[\mathbf{b}(\tau_{1}), \mathbf{b}(\tau_{2}), \dots, \mathbf{b}(\tau_{K})\big] $, $ \mathbf{b}(\tau_{k}) $ $ =  \big[p_{1} \mathbf{a}_{t}^{T}\left(\alpha_{k} \right)  \mathbf{w}_{1}, \quad \, p_{2} e^{-j2\pi \Delta f \tau_{k}} \mathbf{a}_{t}^{T}\left(\alpha_{k} \right)  \mathbf{w}_{2}  , \quad \dots, \quad p_{N} $ $ e^{-j2\pi (N-1) \Delta f \tau_{k}} \mathbf{a}_{t}^{T}\left(\alpha_{k} \right) \mathbf{w}_{N}  \big]^{T} $ and $ \mathbf{V}^{(q)} = [ \mathbf{v}^{q} \left( 1 \right),\mathbf{v}^{q} \left( 2 \right), $ $ \dots, \mathbf{v}^{q} \left( N \right) ] $. Organizing $ \left\lbrace \mathbf{Y}^{(q)}  \right\rbrace _{q=1}^{Q} $ as a big matrix, we have
\vspace{-4pt}
\begin{equation}
	\begin{aligned}
		\mathbf{Y} & = \left[
		\begin{array}{c} \mathbf{Y}^{(1)} \\
			\vdots \\
			\mathbf{Y}^{(Q)}
		\end{array} \right]
		= \left(\mathbf{C} \oplus \mathbf{A}_{R} \right)\mathbf{B}^{T}  + \mathbf{V}
		=  \mathbf{S}  + \mathbf{V},
	\end{aligned}
\vspace{-4pt}
\end{equation}
where $ \mathbf{V} = \big[\mathbf{V}^{(1)T}, \dots, \mathbf{V}^{(Q)T} \big]^{T}$. By vertically stacking the columns of $ \mathbf Y, \mathbf S, \mathbf V $, we can get $\mathbf y=\mathbf s +\mathbf v$.

\section{Peformance Analysis and Optimization for Integrated MIMO-OFDM Signals}\label{3}
\subsection{CRLB of MIMO-OFDM Signals}\label{3.1}
In this subsection, we will analyze the CRLB for targets parameters estimation based on $ \mathbf{y} $. For notational simplicity, we define $ \phi_{k} = \sin{\beta_{k}} $ and $ \theta_{k} = \sin{\alpha_{k}} $. The unknown parameters are grouped as $ \boldsymbol{\gamma} = [\boldsymbol{\varphi}_{1}, \dots, \boldsymbol{\varphi}_{K}, \boldsymbol{\varepsilon}]  $ and $\boldsymbol{u}=[\mathbf{u}_{1}^{T}, \dots,\mathbf{u}_{K}^{T}, \boldsymbol{\varepsilon}]^{T}$ where   $\boldsymbol{\varphi}_{k}=[f_{k}, \tau_{k}, \phi_{k}, \theta_{k}]$, $ \boldsymbol{\varepsilon} = [A_{1}, \dots, A_{K}] $ and $\mathbf{u}_{k}=[v_{k}, r_{k}, \beta_{k}, \alpha_{k}]^{T}$. As $ \mathbf{y} $ is the function of $ \{ \boldsymbol\varphi_{k} \}_{k=1}^{K} $, we first compute the Fisher Information Matrix (FIM) $ \mathbf{J}(\boldsymbol{\gamma}) $ based on $\mathbf y$, and then use the chain rule to calculate
\vspace{-8pt}
\begin{equation}
	\begin{aligned}
		\mathbf{J}(\boldsymbol{u}) = \mathbf{Q} \mathbf{J}(\boldsymbol{\gamma}) \mathbf{Q}^{T},
	\end{aligned}
\vspace{-3pt}
\end{equation}
where $ \mathbf{Q} $ is the Jacobian matrix. $ \mathbf{Q} $ can be calculated as
\begin{equation}
	\begin{aligned}
		\mathbf{Q} = \frac{\partial\boldsymbol{\gamma}}{\partial\boldsymbol{u}} =\left[\begin{array}{llll}
			\boldsymbol{Q}_{1} & \dots & \boldsymbol{0} & \boldsymbol{0} \\
			\vdots & \ddots & \vdots & \vdots \\
			\boldsymbol{0} & \dots & \boldsymbol{Q}_{K} & \boldsymbol{0} \\
			\boldsymbol{0} & \dots & \boldsymbol{0} & \boldsymbol{I}_{1}
		\end{array}\right],
	\end{aligned}
\end{equation}
where
\begin{equation}
	\begin{aligned}
		\boldsymbol{Q}_{k}=\frac{\partial \boldsymbol{\varphi}_{k}}{\partial \boldsymbol{u}_{k}}=\left[\begin{array}{cccc}
			\frac{1}{\lambda} & 0 & 0 & 0\\
			0 & \frac{1}{c} & 0 & 0 \\
			0 & 0 & \cos{\beta_{k}} & 0 \\
			0 & 0 & 0 & \cos{\alpha_{k}}
		\end{array}\right]
	\end{aligned}
\end{equation}
and $ \boldsymbol{I}_{1} $ is a $ K \times K $ unit matrix. Because $ \mathbf{y} $ is a Gaussian vector with mean $ \boldsymbol{\mu} = \mathbf{s} $ and convariance matrix
\begin{equation}
	\begin{aligned}
		\mathbf{E} = \left[\begin{array}{ccc}
			\sigma_{1}^{2}\boldsymbol{I}_{2} & \boldsymbol{0} & \boldsymbol{0} \\
			\boldsymbol{0} & \ddots & \boldsymbol{0} \\
			\boldsymbol{0} & \boldsymbol{0} & \sigma_{N}^{2}\boldsymbol{I}_{2}
		\end{array}\right],
	\end{aligned}
\end{equation}
where $ \sigma^{2}_{n} = p_{n}n_0T $ and $ \boldsymbol{I}_{2} $ is a $ M_{R}Q \times M_{R}Q $ unit matrix, $ \mathbf{J}(\boldsymbol{\gamma}) $ can be caculated as
\begin{equation}
	\begin{aligned}
		\mathbf{J}(\boldsymbol{\gamma})= 2 \Re \left[\frac{\partial^{H} \boldsymbol{\mu}}{\partial \boldsymbol{\gamma}} \mathbf{E}^{-1} \frac{\partial \boldsymbol{\mu}}{\partial \boldsymbol{\gamma}}\right].
	\end{aligned}
\end{equation}
By ignoring the coupling of $ \{ \boldsymbol{\varphi}_{k} \}_{k=1}^{K} $ and $ \boldsymbol{\varepsilon} $, $ \mathbf{J}(\boldsymbol{\gamma}) $ can be written as
\begin{equation}
	\begin{aligned}
		\mathbf{J}(\boldsymbol{\gamma}) = 2 \Re \left[\begin{array}{ccccc}
			\mathbf{F}_{1}^{(1)} & \mathbf{F}_{2}^{(1)} & \dots & \mathbf{F}_{K}^{(1)} & \mathbf{0}\\
			\mathbf{F}_{1}^{(2)} & \mathbf{F}_{2}^{(2)} & \dots & \mathbf{F}_{K}^{(2)} & \mathbf{0} \\
			\vdots & \vdots & \vdots & \vdots & \vdots \\
			\mathbf{F}_{1}^{(K)} & \mathbf{F}_{2}^{(K)} & \dots & \mathbf{F}_{K}^{(K)} & \mathbf{0} \\
			\mathbf{0}& \mathbf{0} & \dots & \mathbf{0} & \mathbf{O}
		\end{array}\right],
	\end{aligned}
\end{equation}
where $ \mathbf{O} $ is a $ K \times K $ matrix and it has no effect on the CRLB of the estimated parameters. In (14), the principal diagonal matrix $ \mathbf{F}_{k}^{(k)} $ is only related to the $k$-th target and has the major impact on the CRLB of $ \mathbf{u}_{k} $. The non-principal diagonal matrices denote the coupling of parameters estimation for different targets. $ \mathbf{F}_{k}^{(k')} $ is a $ 4 \times 4 $ matrix and can be caculated as
\begin{equation}
	\begin{aligned}
		\mathbf{F}_{k}^{(k')} = \left[\frac{\partial^{H} \boldsymbol{\mu}}{\partial \boldsymbol{\varphi}_{k'}} \mathbf{E}^{-1} \frac{\partial \boldsymbol{\mu}}{\partial \boldsymbol{\varphi}_{k}}\right],
	\end{aligned}
\end{equation}
where $ \frac{\partial \boldsymbol{\mu}}{\partial \boldsymbol{\varphi}_{k}} = ( \frac{\partial \boldsymbol{\mu}}{\partial f_{k}}, \frac{\partial \boldsymbol{\mu}}{\partial \tau_{k}}, \frac{\partial \boldsymbol{\mu}}{\partial \phi_{k}}, \frac{\partial \boldsymbol{\mu}}{\partial \theta_{k}}) $ and the $ m_{r}qn $-th element of each column is given by
\begin{equation}
	\begin{aligned}
		& \left[\frac{\partial \boldsymbol{\mu}}{\partial f_{k}}\right]_{m_{r}+(q-1)M_{R}+(n-1)M_{R}Q}\\ & = \frac{p_{n} A_{k}}{\sqrt{M_{R}}}(j2\pi(q-1)T_{s}) e^{j2\pi f_{k}(q-1)T_{s}} e^{-j2\pi (n-1)\Delta f \tau_{k}} \\
		& \times e^{-j\frac{2\pi}{\lambda}(m_{r}-1)d\phi_{k}} \mathbf{a}_{t}^{T}(\alpha_{k})\mathbf{w}_{n}, \\
		& \left[\frac{\partial \boldsymbol{\mu}}{\partial \tau_{k}} \right]_{m_{r}+(q-1)M_{R}+(n-1)M_{R}Q}\\ & =\frac{p_{n} A_{k}}{\sqrt{M_{R}}}(-j2\pi(n-1)\Delta f) e^{j2\pi f_{k}(q-1)T_{s}} e^{-j2\pi (n-1)\Delta f \tau_{k}} \\
		& \times e^{-j\frac{2\pi}{\lambda}(m_{r}-1)d\phi_{k}} \mathbf{a}_{t}^{T}(\alpha_{k})\mathbf{w}_{n}, \\
		& \left[\frac{\partial \boldsymbol{\mu}}{\partial \phi_{k}}\right]_{m_{r}+(q-1)M_{R}+(n-1)M_{R}Q}\\ & = \frac{p_{n} A_{k}}{\sqrt{M_{R}}}(-j\frac{2\pi}{\lambda}(m_{r}-1)d) e^{j2\pi f_{k}(q-1)T_{s}} e^{-j2\pi (n-1)\Delta f \tau_{k}} \\
		& \times  e^{-j\frac{2\pi}{\lambda}(m_{r}-1)d\phi_{k}} \mathbf{a}_{t}^{T}(\alpha_{k})\mathbf{w}_{n}, \\
		& \left[\frac{\partial \boldsymbol{\mu}}{\partial \theta_{k}}\right]_{m_{r}+(q-1)M_{R}+(n-1)M_{R}Q}\\ & = \frac{p_{n} A_{k}}{\sqrt{M_{R}}} e^{j2\pi f_{k}(q-1)T_{s}} e^{-j2\pi (n-1)\Delta f \tau_{k}} e^{-j\frac{2\pi}{\lambda}(m_{r}-1)d\phi_{k}} \\
		& \times \mathbf{a}_{t1}^{T}(\alpha_{k})\mathbf{w}_{n},
	\end{aligned}
\end{equation}
where $ \mathbf{a}_{t1}(\alpha_{k}) = \frac{1}{\sqrt{M_{T}}} \big[ 0, (-j\frac{2\pi}{\lambda}d)e^{-j\frac{2\pi}{\lambda}d \sin \alpha_{k}}, \dots, $ $ \big(-j\frac{2\pi}{\lambda} $ $ (M_{T}-1)d\big)e^{-j\frac{2\pi}{\lambda}\left( M_{T} - 1\right)d \sin \alpha_{k}} \big] ^{T} $. For simplicity, we define $ p_{n}\boldsymbol{\mu}_{nk}^{(i)} = [\frac{\partial \boldsymbol{\mu}}{\partial \boldsymbol{\varphi}_{k}}](((n-1)M_{R}Q + 1): nM_{R}Q, i) $ which is the vector consisting of elements from the $ ((n-1)M_{R}Q + 1) $-th row to the $ nM_{R}Q $-th row in the $ i $-th column of $ \frac{\partial \boldsymbol{\mu}}{\partial \boldsymbol{\varphi}_{k}} $. Therefore, the $ (i, j) $-th element of $ \mathbf{F}_{k}^{(k')} $ is
\begin{equation}
	\begin{aligned}
		\left[\mathbf{F}_{k}^{(k')}\right]_{i,j} & = \left[\frac{\partial \boldsymbol{\mu}}{\partial \boldsymbol{\varphi}_{k'}}\right]_{(:,i)}^{H}   \mathbf{E}^{-1} \left[\frac{\partial \boldsymbol{\mu}}{\partial \boldsymbol{\varphi}_{k}}\right]_{(:,j)}  \\
		& = \sum_{n=1}^{N} \frac{p_{n}}{n_0T} \boldsymbol{\mu}_{nk'}^{(i)H} \boldsymbol{\mu}_{nk}^{(j)},
	\end{aligned}
\end{equation}
where $\mathbf A_{(:,i)}$ denotes the $i$th column of $\mathbf A$. Then (15) can be further expressed as
\begin{equation}
	\begin{aligned}
		\mathbf{F}_{k}^{(k')}= \left[\begin{array}{cccc}
			\mathbf{p}^{T} \mathbf{g}_{k11}^{(k')} & \mathbf{p}^{T} \mathbf{g}_{k12}^{(k')} & \mathbf{p}^{T} \mathbf{g}_{k13}^{(k')} & \mathbf{p}^{T} \mathbf{g}_{k14}^{(k')} \\
			\mathbf{p}^{T} \mathbf{g}_{k21}^{(k')} & \mathbf{p}^{T} \mathbf{g}_{k22}^{(k')} & \mathbf{p}^{T} \mathbf{g}_{k23}^{(k')} & \mathbf{p}^{T} \mathbf{g}_{k24}^{(k')} \\
			\mathbf{p}^{T} \mathbf{g}_{k31}^{(k')} & \mathbf{p}^{T} \mathbf{g}_{k32}^{(k')} & \mathbf{p}^{T} \mathbf{g}_{k33}^{(k')} & \mathbf{p}^{T} \mathbf{g}_{k34}^{(k')} \\
			\mathbf{p}^{T} \mathbf{g}_{k41}^{(k')} & \mathbf{p}^{T} \mathbf{g}_{k42}^{(k')} & \mathbf{p}^{T} \mathbf{g}_{k43}^{(k')} & \mathbf{p}^{T} \mathbf{g}_{k44}^{(k')}
		\end{array}\right],
	\end{aligned}
\end{equation}
where $ \mathbf{p} = \left[p_{1}, \dots, p_{N}\right]^{T} $, and $ \mathbf{g}_{kij}^{(k')} = \frac{1}{n_0T} $ $ [\boldsymbol{\mu}_{1k'}^{(i)H} \boldsymbol{\mu}_{1k}^{(j)}, \dots, $ $  \boldsymbol{\mu}_{Nk'}^{(i)H} \boldsymbol{\mu}_{Nk}^{(j)}]^{T} $. Based on (9)-(14), we can obtain the following FIM for vector $ \boldsymbol{u} $,
\begin{equation}
	\begin{aligned}
		& \mathbf{J}(\boldsymbol{u}) = \left[\begin{array}{ccccc}
			\mathbf{J}_{1}^{(1)} & \mathbf{J}_{2}^{(1)} & \dots
			& \mathbf{J}_{K}^{(1)} & \mathbf{0}\\
			\mathbf{J}_{1}^{(2)}  & \mathbf{J}_{2}^{(2)}  & \dots
			& \mathbf{J}_{K}^{(2)} & \mathbf{0} \\
			\vdots & \vdots & \vdots & \vdots & \vdots \\
			\mathbf{J}_{1}^{(K)} & \mathbf{J}_{2}^{(K)}  & \dots
			& \mathbf{J}_{K}^{(K)} & \mathbf{0} \\
			\mathbf{0}& \mathbf{0} & \dots & \mathbf{0} & \mathbf{O}
		\end{array}\right],
	\end{aligned}
\end{equation}
where $ \mathbf{J}_{k}^{(k')} $ is a 4 $ \times $ 4 matrix with
\begin{equation}
	\begin{aligned}
		\left[\mathbf{J}_{k}^{(k')}\right]_{i,j} = \mathbf{p}^{T} \mathbf{g}_{i,j}^{kk'} + \mathbf{g}_{i,j}^{(kk')H}\mathbf{p}
	\end{aligned}
\end{equation}
and $ \mathbf{g}_{i,j}^{kk'} = [\boldsymbol{Q}_{k'}]_{i,i} \mathbf{g}_{kij}^{(k')} [\boldsymbol{Q}_{k}]_{j,j} $. Let $\boldsymbol{\widehat{u}}=[\mathbf{u}_{1}^{T}, \dots,\mathbf{u}_{K}^{T}]^{T}$. The CRLB matrix of $ \boldsymbol{\widehat{u}} $ is
\begin{equation}
	\begin{aligned}
		C_{\widehat{\boldsymbol{u}}}(\mathbf{p}) = \left[\begin{array}{cccc}
			\mathbf{J}_{1}^{(1)} & \mathbf{J}_{2}^{(1)} & \dots
			& \mathbf{J}_{K}^{(1)} \\
			\mathbf{J}_{1}^{(2)}  & \mathbf{J}_{2}^{(2)}  & \dots
			& \mathbf{J}_{K}^{(2)}  \\
			\vdots & \vdots & \vdots & \vdots  \\
			\mathbf{J}_{1}^{(K)} & \mathbf{J}_{2}^{(K)}  & \dots
			& \mathbf{J}_{K}^{(K)}  \\
		\end{array}\right]^{-1}.
	\end{aligned}
\end{equation}
From (21), the closed-form expression of each principal diagonal element of $ C_{\widehat{\boldsymbol{u}}}(\mathbf{p}) $ is hard to obtain due to the nonzero non-principal diagonal matrices. However, we can get the lower bound of the principal diagonal element $ C_{\widehat{\boldsymbol{u}}}(\mathbf{p}) $ utilizing the property \cite{21}
\begin{equation}
	\begin{aligned}
		\left[\mathbf{J}_{k}^{(k)}\right]_{i,i}^{-1} \leq \left[C_{\widehat{\boldsymbol{u}}}(\mathbf{p})\right]_{4(k-1)+i, 4(k-1)+i}, i = 1, 2, 3, 4.
	\end{aligned}
\end{equation}
When targets are far seperated and the aperture of receive antenna is large, elements in non-principal diagonal matrices have slight impact on the principal diagonal elements of $ C_{\widehat{\boldsymbol{u}}}(\mathbf{p}) $.

\subsection{Optimization of Transmit Powers for Integrated  MIMO-OFDM Signals}\label{3.2}
According to the derived CRLB and transmission rate, we can see that the power vector $\mathbf p$ and beamforming vectors $\{\mathbf w_n\}_{n=1}^N$  will influence both the estimation performance and communication performance. Obviously, maximizing the received SNRs of targets is beneficial to communication and estimation performance. To maximize the received SNR of target $k$, we set $\mathbf w_n =\mathbf a_t^*(\alpha_k), n\in \mathcal N_k$.  We assume the transmitter can estimate the DoAs with the uplink pilot signals from targets so that the optimal transmit beamforming can be implemented.
By using the massive MIMO array at the transmitter, the transmit beams will be narrow enough, such that the transmit steering vectors satisfy $ \left| \mathbf{a}_{t}^{H}( \alpha_{k}) \mathbf{a}_{t}( \alpha_{k'}) \right| \rightarrow 0, \forall \alpha_{k} \neq \alpha_{k'}, k,k' = 1, \dots, K $ \cite{22}.
Therefore, when $\mathbf w_n =\mathbf a_t^*(\alpha_k) $, the transmission rate to the target $k$ is
\vspace{-10pt}
\begin{equation}
	\begin{aligned}
		R_k = \sum_{n\in \mathcal N_k}\log\Big(1+\frac{|A_{k}^{'}|^2 {M_T} p_n T}{n'_0}\Big)
	\end{aligned}
	\vspace{-8pt}
\end{equation}
and (6) can be simplified as
\begin{equation}
	\begin{aligned}
		\mathbf{y}^{q}(n)
		& = A_{k} p_{n} e^{j2\pi f_{k}(q-1)T_{s}} e^{-j2\pi(n-1) \Delta f \tau_{k}} \mathbf{a}_{r}(\beta_{k}) \\
		& + \mathbf{v}^{q}(n),  n \in \mathcal N_k.
	\end{aligned}
\end{equation}

In this case, different subcarriers are used to illuminate the different targets and the interference between different target echoes can be eliminated so that we can get a more concise expression of CRLB. By defining $\boldsymbol{\varphi}_{k}=[f_{k}, \tau_{k}, \phi_{k}]$ and $\mathbf{u}_{k}=[v_{k}, r_{k}, \beta_{k}]^{T}$, the diagonal matrix $ \boldsymbol{Q}_{k} $ of the Jacobian matrix is
\begin{equation}
	\begin{aligned}
		\boldsymbol{Q}_{k}=\frac{\partial \boldsymbol{\varphi}_{k}}{\partial \boldsymbol{u}_{k}}=\left[\begin{array}{ccc}
			\frac{1}{\lambda} & 0 & 0 \\
			0 & \frac{1}{c} & 0 \\
			0 & 0 & \cos{\beta_{k}}
		\end{array}\right].
	\end{aligned}
\end{equation}
On the other hand, in (15), $ \mathbf{F}_{k}^{(k)} $s are simplified as $ 3 \times 3 $ diagonal matrices and  the $ \mathbf{F}_{k}^{(k')} = \mathbf{0} $ for $ k \neq k'$ . Define $ \frac{\partial \boldsymbol{\mu}}{\partial \boldsymbol{\varphi}_{k}} = ( \frac{\partial \boldsymbol{\mu}}{\partial f_{k}}, \frac{\partial \boldsymbol{\mu}}{\partial \tau_{k}}, \frac{\partial \boldsymbol{\mu}}{\partial \phi_{k}}) $, where the $ m_{r}qn $-th element can be written as
\begin{equation}
	\begin{aligned}
		& \left[\frac{\partial \boldsymbol{\mu}}{\partial f_{k}}\right]_{m_{r}+(q-1)M_{R}+(n-1)M_{R}Q} \\ & = \left\lbrace
		\begin{aligned}
			& \frac{p_{n} A_{k}}{\sqrt{M_{R}}}(j2\pi(q-1)T_{s}) e^{j2\pi f_{k}(q-1)T_{s}} \\
			& \times e^{-j2\pi (n-1)\Delta f \tau_{k}} e^{-j\frac{2\pi}{\lambda}(m_{r}-1)d\phi_{k}}, n \in \mathcal N_k \\
			& 0, \qquad\qquad\qquad\qquad\qquad\qquad\qquad  n \notin \mathcal N_k
		\end{aligned} \right. \\
		& \left[\frac{\partial \boldsymbol{\mu}}{\partial \tau_{k}}\right]_{m_{r}+(q-1)M_{R}+(n-1)M_{R}Q} \\ & = \left\lbrace
		\begin{aligned}
			& \frac{p_{n} A_{k}}{\sqrt{M_{R}}}(-j2\pi(n-1)\Delta f) e^{j2\pi f_{k}(q-1)T_{s}} \\
			& \times e^{-j2\pi (n-1)\Delta f \tau_{k}} e^{-j\frac{2\pi}{\lambda}(m_{r}-1)d\phi_{k}}, n \in \mathcal N_k \\
			& 0, \qquad\qquad\qquad\qquad\qquad\qquad\qquad  n \notin \mathcal N_k
		\end{aligned} \right. \\ 		
		& \left[\frac{\partial \boldsymbol{\mu}}{\partial \phi_{k}}\right]_{m_{r}+(q-1)M_{R}+(n-1)M_{R}Q} \\ & = \left\lbrace
		\begin{aligned}
			& \frac{p_{n} A_{k}}{\sqrt{M_{R}}}(-j\frac{2\pi}{\lambda}(m_{r}-1)d) e^{j2\pi f_{k}(q-1)T_{s}} \\
			& \times e^{-j2\pi (n-1)\Delta f \tau_{k}} e^{-j\frac{2\pi}{\lambda}(m_{r}-1)d\phi_{k}}, n \in \mathcal N_k \\
			& 0, \qquad\qquad\qquad\qquad\qquad\qquad\qquad  n \notin \mathcal N_k
		\end{aligned} \right. . \\
	\end{aligned}
\end{equation}
Based on the above analysis, $ \mathbf{J}(\boldsymbol{u}) $ in this case will degenerate to the following block diagonal matrix
\begin{equation}
	\begin{aligned}
		& \mathbf{J}(\boldsymbol{u}) = \left[\begin{array}{ccccc}
			\mathbf{J}_{1}^{(1)} & \mathbf{0} & \dots
			& \mathbf{0} & \mathbf{0}\\
			\mathbf{0}  & \mathbf{J}_{2}^{(2)}  & \dots
			& \mathbf{0} & \mathbf{0} \\
			\vdots & \vdots & \vdots & \vdots & \vdots \\
			\mathbf{0} & \mathbf{0} & \dots
			& \mathbf{J}_{K}^{(K)} & \mathbf{0} \\
			\mathbf{0}& \mathbf{0} & \dots & \mathbf{0} & \mathbf{O}
		\end{array}\right],
	\end{aligned}
\end{equation}
where $ \mathbf{J}_{k}^{(k)} $ is a 3 $ \times $ 3 matrix and can be calculated with (20). Therefore, the CRLB matrix of the $ k $-th target is
\vspace{0pt}
\begin{equation}
	\begin{aligned}
		\mathbf C_{k}(\mathbf{p}) = [\mathbf{J}_{k}^{(k)}]^{-1} =  \frac{1}{2}\left[\begin{array}{ccc}
			\mathbf{p}^{T} \mathbf{g}_{1,1}^{kk} & \mathbf{p}^{T} \mathbf{g}_{1,2}^{kk} & \mathbf{p}^{T} \mathbf{g}_{1,3}^{kk} \\
			\mathbf{p}^{T} \mathbf{g}_{2,1}^{kk} & \mathbf{p}^{T} \mathbf{g}_{2,2}^{kk} & \mathbf{p}^{T} \mathbf{g}_{2,3}^{kk}\\
			\mathbf{p}^{T} \mathbf{g}_{3,1}^{kk} & \mathbf{p}^{T} \mathbf{g}_{3,2}^{kk} & \mathbf{p}^{T} \mathbf{g}_{3,3}^{kk}\\
		\end{array}\right]^{-1},
	\end{aligned}
\end{equation}
where $ [\mathbf g_{1,1}^{kk}]_n=\frac{ (A_k 2 \pi T_s)^2 Q(Q-1)(2Q-1) }{6\lambda^2 n_0T}$,
$ [\mathbf g_{1,2}^{kk}]_n = [\mathbf g_{2,1}^{kk}]_n $ $  =  \frac{-(A_k2\pi)^2T_s\Delta f Q(Q-1)(2Q-1) (n-1)}{6c\lambda n_0 T}$,
$ [\mathbf g_{2,2}^{kk}]_n = $$ \frac{Q(A_k2\pi(n-1)\Delta f)^2}{c^2 n_0 T}$,
$ [\mathbf g_{1,3}^{kk}]_n = [\mathbf g_{3,1}^{kk}]_n $ $  =  \frac{ -(A_k2\pi)^2 dT_s QM_R(M_R-1)(Q-1)\cos \beta_k}{\lambda^2 n_0 T}$,
$ [\mathbf g_{2,3}^{kk}]_n = [\mathbf g_{3,2}^{kk}]_n = \frac{ (A_k2\pi)^2 \Delta f d Q (M_R-1) (n-1)\cos\beta_k }{2c\lambda n_0T }$,
$ [\mathbf g_{3,3}^{kk}]_n = \frac{  Q (A_k2\pi d)^2 (M_R-1)(2M_R-1) \cos^2 \beta_k }{6\lambda^2 n_0T }$ for
$ n \in \mathcal N_k $ and the other items with $n \notin \mathcal N_k$ equal 0.

Since the inverse matrix of matrix $ \mathbf{J}_{k}^{(k)} $ is complicated for further analysis, we use the following approximate method to obtain a lower bound of CRLB. By ignoring elements in non-principal diagonal blocks of $\mathbf J_k^{(k)}$, the CRLB matrices of unknown parameters of the $k$th target are given by

\begin{equation}\label{eq30}
	\begin{aligned}
		\mathbf C_{k}^{vr}(\mathbf{p}) = \frac{1}{2}\left[\begin{array}{cc}
			\mathbf{p}^{T} \mathbf{g}_{1,1}^{kk} & \mathbf{p}^{T} \mathbf{g}_{1,2}^{kk}\\
			\mathbf{p}^{T} \mathbf{g}_{2,1}^{kk} & \mathbf{p}^{T} \mathbf{g}_{2,2}^{kk}\\
		\end{array}\right]^{-1}
	\end{aligned}
\end{equation}
and
\begin{equation}\label{eq31}
	\begin{aligned}
		C_{k}^{\beta}(\mathbf{p}) = \left(2\mathbf{p}^{T} \mathbf{g}_{3,3}^{kk}\right)^{-1}.
	\end{aligned}
\end{equation}
Based on \eqref{eq30} and \eqref{eq31}, the lower bound of CRLB of the estimated parameters for the $k$th target can be caculated as
\vspace{-3pt}
\begin{align}
	LCRLB_{v_{k}} = \frac{\mathbf{p}^{T} \mathbf{g}_{2,2}^{kk}}{\mathbf{p}^{T}\mathbf{G}_{k}\mathbf{p}}, \label{eq32}\\
	LCRLB_{r_{k}} = \frac{\mathbf{p}^{T} \mathbf{g}_{1,1}^{kk}}{\mathbf{p}^{T}\mathbf{G}_{k}\mathbf{p}},\label{eq33}\\
	LCRLB_{\beta_{k}} = \frac{1}{2\mathbf{p}^{T} \mathbf{g}_{3,3}^{kk}},\label{eq34}
\end{align}
where
$ \mathbf{G}_{k} = 2 \Big[\mathbf{g}_{1,1}^{kk} \mathbf{g}_{2,2}^{(kk)T}  - \mathbf{g}_{1,2}^{kk} \mathbf{g}_{2,1}^{(kk)T} \Big]
$. It can be seen that $LCRLB_{\beta_k}$ in \eqref{eq34} is only related to the total power allocated to subcarrier set $\mathcal N_k$ and has nothing to do with the power distribution within the subcarrier set $\mathcal N_k$.

Since the power allocation among subcarriers influences both estimation and communication performances, the power allocation should be optimized by considering the tradeoff between transmission rate and CRLB. The problem is formulated as
\begin{align}
	&\min_{\mathbf{p}} \quad -\sum_{k=1}^{K} \sum_{n\in \mathcal N_k}\log \Big(1+\frac{|A_{k}^{'}|^2 {M_T} p_n T}{n'_0}\Big) \label{optiz-problem} \\
	&\; \text{s.t.} \qquad \sum_{n=1}^{N} p_{n} \leq p_{T}, \tag{\ref{optiz-problem}a} \\
	&\qquad \; LCRLB_{v_{k}} \leq \eta_{k}^{v}, \forall k, \tag{\ref{optiz-problem}b} \\
	&\qquad \; LCRLB_{r_{k}} \leq \eta_{k}^{r}, \forall k, \tag{\ref{optiz-problem}c}\\	
	&\qquad \; LCRLB_{\beta_{k}} \leq \eta_{k}^{\beta}, \forall k, \tag{\ref{optiz-problem}d}
\end{align}
where $ p_{T} $ is the total power constraint and $ \eta_{k}^{v}, \eta_{k}^{r}, \eta_{k}^{\beta} $ are the CRLB constraints of speed, distance, and DoA estimations respectively.

It seems that the CRLB constraints in ({\ref{optiz-problem}b}) and ({\ref{optiz-problem}c}) are not convex functions of $\mathbf p$. However, we can transform these constraints into convex ones. For convenience, we define $ \left[\mathbf{J}_{k}^{(k)}\right]_{i,j} = c_{ij} $. Since $ LCRLB_{v_{k}} = [\mathbf C_{k}^{vr}(\mathbf{p})]_{1,1} $, we have
\begin{equation}
	\begin{aligned}
		\left[\begin{array}{cc}
			c_{11}  & c_{12}\\
			c_{21} & c_{22}\\
		\end{array}\right]^{-1}_{1,1}
		&\leq \eta_{k}^{v} \Rightarrow \frac{{c}_{22}}{{c}_{11}{c}_{22}-{c}_{12}{c}_{21}} \leq \eta_{k}^{v} \\
		&\Rightarrow {c}_{22} ({c}_{11} \eta_{k}^{v} - 1) - {c}_{12}{c}_{21}\eta_{k}^{v} \geq 0 \: \:.
	\end{aligned}
\end{equation}
By constructing the matrix
\begin{equation}
	\begin{aligned}
		\widetilde{\mathbf{C}}^{v}_{k} = \left[\begin{array}{cc}
			c_{22}  & c_{12} \sqrt{\eta_{k}^{v}}\\
			c_{21}\sqrt{\eta_{k}^{v}} & c_{11}\eta_{k}^{v}-1\\
		\end{array}\right],
	\end{aligned}
\end{equation}
we can see that the $ \widetilde{\mathbf{C}}^{v}_{k} $ is a positive definite matrix, i.e., $ 		LCRLB_{v_{k}} \leq \eta_{k}^{v} \Longleftrightarrow \widetilde{\mathbf{C}}^{v}_{k} \succeq 0. $
Similarly, the constraint in ({\ref{optiz-problem}c}) can be transformed into $\widetilde{\mathbf C}_k^r \succeq 0$. Based on the above analysis, the optimization problem in ({\ref{optiz-problem}}) can be re-expressed as
\begin{align}
	&\min_{\mathbf p} \quad -\sum_{k=1}^{K} \sum_{n\in \mathcal N_k}\log \Big(1+\frac{|A_{k}^{'}|^2 {M_T} p_n T}{n'_0}\Big) \label{optimize-problem}\\
	&\; \text{s.t.} \qquad \sum_{n=1}^{N} p_{n} \leq p_{T},  \tag{\ref{optimize-problem}a} \\
	&\qquad \: \widetilde{\mathbf{C}}^{v}_{k} \succeq 0, \widetilde{\mathbf{C}}^{r}_{k} \succeq 0, \forall k, \tag{\ref{optimize-problem}b} \\ 	
	&\qquad \; LCRLB_{\beta_{k}} \leq \eta_{k}^{\beta}, \forall k, \tag{\ref{optimize-problem}c}
\end{align}
where
\begin{align}
	\widetilde{\mathbf{C}}^{v}_{k} = \left[\begin{array}{cc}
		2\mathbf{p}^{T} \mathbf{g}_{2,2}^{kk} & 2\mathbf{p}^{T} \mathbf{g}_{1,2}^{kk}\sqrt{\eta_{k}^{v}}\\
		2\mathbf{p}^{T} \mathbf{g}_{2,1}^{kk}\sqrt{\eta_{k}^{v}} & 2\mathbf{p}^{T} \mathbf{g}_{1,1}^{kk} \eta_{k}^{v}-1\\
	\end{array}\right],\\
	\widetilde{\mathbf{C}}^{r}_{k} = \left[\begin{array}{cc}
		2\mathbf{p}^{T} \mathbf{g}_{1,1}^{kk}  & 2\mathbf{p}^{T} \mathbf{g}_{1,2}^{kk} \sqrt{\eta_{k}^{r}} \\
		2\mathbf{p}^{T} \mathbf{g}_{2,1}^{kk}\sqrt{\eta_{k}^{r}} & 2\mathbf{p}^{T} \mathbf{g}_{2,2}^{kk} \eta_{k}^{r}-1 \\
	\end{array}\right].
\end{align}
The optimization problem in $ {\eqref{optimize-problem}} $ is a semidefinite programming (SDP) and can be solved by CVX toolbox. Moreover, for given Lagrangian multiplies, we can get the closed-form expression of transmit powers, which can provide some insights for power allocation method in integrated MIMO-OFDM systems.
The Lagrangian associated with problem in $ {\eqref{optimize-problem}} $ is
\vspace{0pt}
\begin{equation}
	\begin{aligned}
		\mathcal{L}
		 =& -\sum_{k=1}^{K} \sum_{n\in \mathcal  N_k}\log\Big(1+\frac{|A_{k}^{'}|^2 {M_T} p_n T}{n'_0}\Big)\\
		 &+ \lambda_{1} \Big(\sum_{n=1}^{N} p_{n}
		 - p_{T}\Big)  +\sum_{k=1}^{K}\mu_{k}\Big(\frac{1}{\eta_{k}^{\beta}} - 2\mathbf{p}^{T} \mathbf{g}_{3,3}^{kk} \Big)\\
		 &-\sum_{k=1}^{K}\text{tr}(\mathbf{Z}_{k}^{v}\widetilde{\mathbf{C}}^{v}_{k}-\sum_{k=1}^{K}\text{tr}(\mathbf{Z}_{k}^{r}\widetilde{\mathbf{C}}^{r}_{k})
	\end{aligned}
\end{equation}
where $ \lambda_{1} \geq 0, \mu_{k} \geq 0, \mathbf{Z}_{k}^{v} \succeq 0,
\mathbf{Z}_{k}^{r} \succeq 0 $
are Lagrangian multiplies.
The partial derivative of $\mathcal L$ with respect to $p_n$ is
\begin{equation}
	\begin{aligned}
		\frac{\partial \mathcal{L}}{\partial p_{n}}
		=& \frac{-|A_{k}^{'}|^2 {M_T} T}{(n'_0 + |A_{k}^{'}|^2 {M_T} p_n T) \ln2} + \lambda_{1} - \sum_{k=1}^{K} 2\mu_{k}\mathbf{g}_{3,3}^{kk}(n)\\ & -\sum_{k=1}^{K}2\text{tr}(\mathbf{Z}_{k}^{v}\mathbf{L}(\eta_{k}^{v})) -\sum_{k=1}^{K}2\text{tr}(\mathbf{Z}_{k}^{r}\mathbf{L}(\eta_{k}^{r})), \forall n,
	\end{aligned}
\end{equation}
where
\vspace{0pt}
\begin{align}
	\mathbf{L}(\eta_{k}^{v}) = \left[\begin{array}{cc}
		[\mathbf{g}_{2,2}^{kk}]_n&[\mathbf{g}_{1,2}^{kk}]_n\sqrt{\eta_{k}^{v}} \\
		\left[\mathbf{g}_{2,1}^{kk}\right]_n\sqrt{\eta_{k}^{v}}&[\mathbf{g}_{1,1}^{kk}]_n\eta_{k}^{v} \\
	\end{array}\right],\\
	\mathbf{L}(\eta_{k}^{r}) = \left[\begin{array}{cc}
		[\mathbf{g}_{1,1}^{kk}]_n&[\mathbf{g}_{1,2}^{kk}]_n\sqrt{\eta_{k}^{r}} \\
		\left[\mathbf{g}_{2,1}^{kk}\right]_n\sqrt{\eta_{k}^{r}}&[\mathbf{g}_{2,2}^{kk}]_n\eta_{k}^{r} \\
	\end{array}\right].
\end{align}
By letting $ \frac{\partial \mathcal{L}}{\partial p_{n}} = 0 $, the optimal transmit power to minimize $\mathcal L$ is given by
\begin{equation}\label{eq45}
	\begin{aligned}
		p_n= \Big[\frac{1}{(\lambda_{1} - \xi)\ln2} - \frac{n'_0}{|A_{k}^{'}|^2 {M_T} T} \Big]^+, \forall n,
	\end{aligned}
\end{equation}
where $ [x]^+ = \max(x, 0)$ and
\begin{equation}
	\begin{aligned}
		\xi &= \sum_{k=1}^{K} 2\Big(\mu_{k}[\mathbf{g}_{3,3}^{kk}]_n+ \text{tr}(\mathbf{Z}_{k}^{v}\mathbf{L}(\eta_{k}^{v}))+
		\text{tr}(\mathbf{Z}_{k}^{r}\mathbf{L}(\eta_{k}^{r}))\Big).
	\end{aligned}
\end{equation}
From \eqref{eq45}, we can see that the power allocation has the ``water-filling" form, and the estimation performance constraints just modify the water level.
 
\section{Multi-target estimation based on MIMO-OFDM signals}
\label{4}
\subsection{Serial-Parallel (SP) Estimation}\label{4.1}

According to the traditional serial-parallel estimation, we first estimate DoA, and then estimate another two parameters jointly with 2D-FFT MUSIC or compressed sensing in \cite{2010Signal}.

Using received data $ \mathbf{Y} $ in (8), the DoA of targets can be estimated by typical MUSIC algorithm. Let $ \widehat{\mathbf{A}}_{R} = [\mathbf a_r(\widehat{\beta}_1), \dots, \mathbf a_r(\widehat{\beta}_K)] $ denote the receiving array matrix, where $\widehat{\beta}_1, \dots , \widehat{\beta}_K$ are estimated DoAs. Multiplying (7) by the pseudo inverse of $\widehat{\mathbf{A}}_{R}$ ($K \leq M_R$), we get
\begin{equation}
	\begin{aligned}
		\widehat{\mathbf{A}}_{R}^{\dag} \mathbf{Y}^{(q)} & = \widehat{\mathbf{A}}_{R}^{\dag} \mathbf{A}_{R} D_{q}\left( \mathbf{C} \right)
		\mathbf{B}^{T} + \widehat{\mathbf{A}}_{R}^{\dag} \mathbf{V}^{(q)} \\ & = \mathbf A D_{q}\left( \mathbf{C} \right)
		\mathbf{B}^{T} + \widehat{\mathbf{A}}_{R}^{\dag}\mathbf{V}^{(q)}
	\end{aligned}
\end{equation}
where $\mathbf A$ is a diagonally dominant matrix. Adding up each row of $\widehat{\mathbf{A}}_R^{\dag} \mathbf Y^q$, we have
\vspace{-0.2cm}
\begin{equation}\label{eq48}
	\begin{aligned}
		\widehat{\mathbf{y}}^q &= \sum_{k=1}^{K}
		A_k B_k^{'} e^{j2\pi f_k (q-1)T_s} \mathbf a_t^T(\alpha_{k}) \\
		& \times [\mathbf w_1p_1, \dots, \mathbf w_N p_N e^{-j2\pi\left( N-1\right) \Delta f \tau_{k}}] +\mathbf u^q
	\end{aligned}
\end{equation}
where $ B_k^{'} = \mathbf 1^T (\widehat{\mathbf{A}}_R^{\dag} \mathbf a_r(\beta_k))$, $ \mathbf u^q =\mathbf 1^T (\widehat{\mathbf{A}}_R^{\dag}  \mathbf V^{(q)})$ and all elements of $\mathbf 1$ are 1. Because subcarriers are allocated to $ K $ targets, the vectors in \eqref{eq48} are divided into $K$ groups and the same transmit beamforming vector is used in each group. The data in the $k$-th group of $\widehat{\mathbf{y}}^q$ is used to estimate the speed and distance of the $k$-th target. Due to the estimation error of the DoAs, the estimation of the $k$-th target will be interfered by the echo signals of other targets. We assume that when $n\in \mathcal N_k$, $\mathbf w_n = \mathbf a_t^*(\alpha_k)$. Then, the signal model for estimating $(f_k, \tau_k)$ is
\begin{equation}
	\begin{aligned}
		\widehat y^{q}(n) &= \sum_{k=1}^{K} A_k B_{k,n} p_n e^{j2\pi f_{k}(q-1)T_{s}}
		e^{-j2\pi\left( n-1\right) \Delta f \tau_{k}} \\ &  + u^{q}(n),   n \in \mathcal N_k,
	\end{aligned}
\end{equation}
where $ B_{k,n} = B_k^{'} \mathbf a_t^T(\alpha_{k}) \mathbf w_n $. With $\{\widehat y^q(n), q=1,\cdots, Q, n\in \mathcal N_k\}$, the velocity and range of the $k$th target can be estimated with spatial smoothing two-dimensional MUSIC (2D-MUSIC) or compressed sensing \cite{2010Signal}.

\begin{figure}
	\centerline{\includegraphics[scale=0.015]{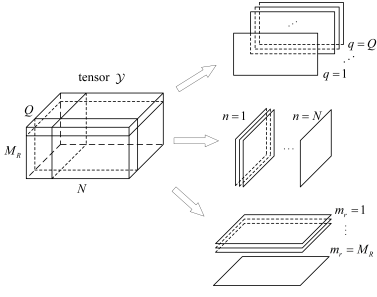}}
	\caption{Unfold third-order echo signal tensor with frontal slabs, lateral slabs and horizontal slabs respectively.}
	\label{fig2}
	\vspace{-15pt}
\end{figure}
\subsection{Estimation Based on Tensor Decomposition}\label{4.2}
In the traditional SP estimation, angel estimation error will effect the estimation accuracy of velocity and distance. To decouple the estimation of angle, velocity and distance, we propose the estimation method based on tensor decomposition. According to the multi-dimensional signal processing model in \cite{23}, we express the received signal model in (8) in the form of the following third-order tensor
\begin{align}\label{eq50}
	\mathcal Y = \mathcal S + \mathcal V
\end{align}
where $\mathcal S = \sum_{k=1}^K \mathbf a_r(\beta_k) \circ \mathbf b(\tau_k) \circ \mathbf c(f_k)$.  Comparing \eqref{eq50} with (6), we can see that the element of $\mathcal S$ at $(m,n,q)$ is $ {\mathcal S}_{mnq}= \sum_{k=1}^{K} \frac{A_{k}}{\sqrt{M_R}}p_n e^{j2\pi f_{k}(q-1)T_{s}} e^{-j2\pi(n-1) \Delta f \tau_{k}} e^{-j\frac{2\pi}{\lambda}(m_{r}-1)d\phi_{k}} \times \mathbf{a}_{t}^{T}(\alpha_{k}) \mathbf{w}_{n}$.  It is known that outer product of three vectors is a rank-1 third-order tensor. From \eqref{eq50}, we can see that $\mathcal S$ is the sum of $K$ rank-1 tensors. In the tensor decompositions, decomposing a tensor into several rank-1 tensors is called canonical polyadic decomposition (CPD) \cite{23}. Therefore, our objective is to obtain CPD of $\mathcal S$ from the noisy observation $\mathcal Y \in C^{M_R\times N \times Q}$ so that the difference between $\mathcal Y$ and $\mathcal S$ is minimized, i.e,
\begin{align}\label{eq51}
	\min_{\{\mathbf a_r(\beta_k), \mathbf b(\tau_k), \mathbf c(f_k)\}_{k=1}^K} \|\mathcal Y-\mathcal S\|.
\end{align}
To solve the problem \eqref{eq51}, we first define  $ \mathbf{A}_R \in \mathbb{C}^{M_R \times K} $, $ \mathbf{B} \in \mathbb{C}^{N \times K}$, $ \mathbf{C} \in \mathbb{C}^{Q \times K} $ as factor matrices of $\mathcal S$. As shown in Fig.\ref{fig2}, we first consider the frontal slab of $\mathcal S$, $ \mathbf{S}^{q} = \mathbf{A}_R D_{q}\left( \mathbf{C} \right) \mathbf{B}^{T} $.
By vertical stacking $ \mathbf{S}^{q} $, we get the unfolding matrix $ \mathbf{S}_{1} = \left(\mathbf{C} \oplus \mathbf{A}_R \right)\mathbf{B}^{T} $.
In the same way, the lateral and horizontal slabs of $\mathcal S$ are expressed as $ \mathbf{S}^{n} = \mathbf{C} D_{n}\left( \mathbf{B} \right) \mathbf{A}_R^{T}$ and $ \mathbf{S}^{m_r} = \mathbf{B} D_{m_r}\left( \mathbf{A}_R \right) \mathbf{C}^{T} $.

By unfolding the frontal, lateral and horizontal slabs of $ \mathcal{Y} $, we can get the following expressions $ \mathbf{Y}_{1} = \mathbf{S}_{1} + \mathbf{V}_{1}$, $  \mathbf{Y}_{2} = \mathbf{S}_{2} + \mathbf{V}_{2}$, $
\mathbf{Y}_{3} = \mathbf{S}_{3} + \mathbf{V}_{3} $,
where $ \mathbf{S}_{2} = \left( \mathbf{B} \oplus \mathbf{C} \right) \mathbf{A}_R^{T} $, $ \mathbf{S}_{3} = \left( \mathbf{A}_R \oplus \mathbf{B} \right) \mathbf{C}^{T} $, $ \{\mathbf{V}_{v}\}_{v = 1}^{3} $ are the noise terms and $ \mathbf{Y}_{1} \in \mathbb{C}^{Q M_{R}\times N  } $, $ \mathbf{Y}_{2} \in \mathbb{C}^{N Q\times M_{R} } $, $ \mathbf{Y}_{3} \in \mathbb{C}^{ M_{R}N \times Q} $ are the unfolding matrices of $ \mathcal{Y} $ by different modes. Suppose that the number of the targets is known. With unfolding matrices, the problem in \eqref{eq51} can be re-expressed as
\begin{equation}
	\begin{aligned}	
		\min_{\mathbf{A}_{R}, \mathbf{B}, \mathbf{C}}\left| \left| \mathbf{Y}_{1} - \left(\mathbf{C} \oplus \mathbf{A}_{R} \right)\mathbf{B}^{T} \right| \right|_{F}^{2}.
	\end{aligned}
\end{equation}

The above problem can be solved via $ \textit{Alternating Least Squares} \mathbf{(ALS)} $ algorithm \cite{23}, which alternately optimizes $\mathbf A_R, \mathbf B, \mathbf C$ utilizing the following update equations
\vspace{0pt}
\begin{equation}
	\begin{aligned}	
		\widehat{\mathbf{B}} \leftarrow \arg \min_{\mathbf{B}}\left| \left| \mathbf{Y}_{1} - \left(\widehat{\mathbf{C}} \oplus \widehat{\mathbf{A}}_{R} \right)\mathbf{B}^{T} \right| \right|_{F}^{2},
	\end{aligned}
\end{equation}
\begin{equation}
	\begin{aligned}	
		\widehat{\mathbf{A}}_{R} \leftarrow \arg \min_{\mathbf{A}_{R}}\left| \left| \mathbf{Y}_{2} - \left( \widehat{\mathbf{B}} \oplus \widehat{\mathbf{C}} \right) \mathbf{A}_{R}^{T} \right| \right|_{F}^{2},
	\end{aligned}
\end{equation}
\begin{equation}
	\begin{aligned}	
		\widehat{\mathbf{C}} \leftarrow \arg \min_{\mathbf{C}}\left| \left| \mathbf{Y}_{3} - \left( \widehat{\mathbf{A}}_{R} \oplus \widehat{\mathbf{B}} \right) \mathbf{C}^{T} \right| \right|_{F}^{2},
	\end{aligned}
\end{equation}
until convergence.

Next, we discuss the uniqueness of the CPD.

\textbf{Theorem \cite{24}:} For a third-order tensor $ \mathcal{X} \in \mathbb{C}^{I_1 \times I_2 \times I_3} $ with factor matrices $ \mathbf A_1 \in \mathbb{C}^{I_1 \times K} $, $ \mathbf A_2 \in \mathbb{C}^{I_2 \times K} $, $ \mathbf A_3 \in \mathbb{C}^{I_3 \times K}$, the uniqueness condition of the CPD is
\begin{align}\label{eq60}
	k_{\mathbf A_1} + k_{\mathbf A_2} + k_{\mathbf A_3} \geq 2K+2,
\end{align}
where $ k_{\mathbf{U}} $ is the Kruskal rank of $ \mathbf{U} $, which is defined as the largest integer $ k $ that any $ k $ columns of the $ \mathbf{U} $ are linearly independence.

\textbf{Proof:} The detailed proof is given in \cite{24}.

Note that the original proof of Kruskal's condition is not intuitive enough, and the intuitive proof can be found in \cite{25}. Also, it does not mean the CPD is not unique beyond the Kruskal's condition \cite{23}.

From the above theorem, we know that if $k_{\mathbf A_R} + k_\mathbf B + k_\mathbf C \geq 2K+2$,
the CPD of $ \mathcal Y $ is unique. We first examine the Kruskal rank of $ \mathbf A_R $ and $ \mathbf C $. Since $ \mathbf A_R $ and $ \mathbf C $ are column-wise scaled Vandermonde matrix respectively, the Kruskal rank of $ \mathbf A_R $ and $ \mathbf C $ are $ k_{\mathbf A_R} = \min(M_R, K) $ and $ k_{\mathbf C} = \min(Q, K) $.
Then let us consider the Kruskal rank of $ \mathbf B $. As stated in \ref{3.2}, we set $\mathbf w_n =\mathbf a_t^*(\alpha_k)$ when $n\in \mathcal N_k$. According to the definition of $\mathbf b(\tau_k)$ in \ref{2} and the analysis in \ref{3.2}, terms associated with $\mathbf a_t^T(\alpha_k) \mathbf w_n, n \notin \mathcal N_k$ in $ \mathbf b(\tau_k)$ will vanish with the increasing of $M_T$. Therefore, we can always find a $ K\times K $ diagonally dominant submatrix of $ \mathbf B$ when the subcarriers are assigned to $ K $ targets and $ N \geq K $. The diagonally dominant matrix is invertible which means $ \mathbf B $ is full column rank, so $ k_{\mathbf B} = rank (\mathbf B) = K $.
When $ K \leq \max(M_R, Q, N) $, we have $ k_{\mathbf A_R} + k_{\mathbf B} + k_{\mathbf C} = 3K $. When $ K \geq 2 $, the \eqref{eq60} can be satisfied.
The uniqueness holds for $ K = 1 $, irrespective of condition \eqref{eq60}, as
long as $ \mathcal{S}$ does not contain an identically zero slab along
any dimension \cite{26}.

It is well known that the CPD is unique up to inherent permutation ambiguity and scale ambiguity, i.e., $ \widehat{\mathbf{A}}_{R} = \mathbf{A}_R \mathbf\Pi \mathbf\Lambda_a$, $ \widehat{\mathbf{B}} = \mathbf{B} \mathbf\Pi \mathbf\Lambda_b$, $ \widehat{\mathbf{C}} = \mathbf{C} \mathbf\Pi \mathbf\Lambda_c$, where $ \mathbf\Pi $ is a permutation matrix and $\mathbf\Lambda_a $, $ \mathbf\Lambda_b $, $ \mathbf\Lambda_c $ are diagonal matrices which satisfied $ \mathbf\Lambda_a \mathbf\Lambda_b \mathbf\Lambda_c = \mathbf I_K $.
It is worth noting that permutation ambiguity and scale ambiguity of CPD will not affect the subsequent estimation of parameters. Firstly, the permutation matrix is the same which means the permutation ambiguity does not change the correlation of the parameters for the same column of the factor matrices $ \widehat{\mathbf{A}}_{R} $, $ \widehat{\mathbf{B}} $, $ \widehat{\mathbf{C}} $. Secondly, we mainly utilize the phase characteristics of $ \mathbf a_r(\beta_k) $, $ \mathbf b(\tau_k) $ and $ \mathbf c(f_k) $ to estimate the parameters. Hence both of them have no influence on the parameter estimation.

After CPD of the receive echo signal, we obtain the estimated values of the factor matrices $ \widehat{\mathbf{A}}_{R} = [\widehat{\mathbf a}_r(\beta_1), \dots, \widehat{\mathbf a}_r(\beta_K)] $, $ \widehat{\mathbf{B}} = [\widehat{\mathbf b}(\tau_1), \dots, \widehat{\mathbf b}(\tau_K)] $, $ \widehat{\mathbf{C}} = [\widehat{\mathbf c}(f_1), \dots, \widehat{\mathbf c}(f_K)] $. We can see that the same column of the factor matrices corresponds to parameters $ \beta_{k} $, $ \tau_{k} $ and $ f_{k} $  to be estimated for the same target. To estimate $\beta_k$  from $ \widehat{\mathbf a}_r(\beta_k) $, inverse discrete Fourier transform (IDFT) is conducted for $ \widehat{\mathbf a}_r(\beta_k) $
\begin{equation}
	\begin{aligned}	
		ang(l_{\beta_{k}}) & = {\rm IDFT} [\widehat{\mathbf a}_r(\beta_k)]  \\
		& = \dfrac{1}{M_{R}} \sum _{m_{r} = 1}^{M_{R}} a_{k}(m_{r}) \exp \left(j \frac{2 \pi}{M_{R}}(m_{r}-1)l_{\beta_{k}}\right),
	\end{aligned}
\end{equation}
where $a_k(m_r)$ is the $m_r$th element of $\widehat{\mathbf a}_r(\beta_k)$. Also, discrete Fourier transform (DFT) is applied to $ \widehat{\mathbf c}(f_k) $
\begin{equation}
	\begin{aligned}	
		vel(l_{v_{k}}) & = {\rm DFT} [\widehat{\mathbf c}(f_k)]  \\
		& = \dfrac{1}{Q} \sum _{q = 1}^{Q} c_{k}(q) \exp \left(-j\frac{2 \pi}{Q}(q-1)l_{v_{k}}\right),
	\end{aligned}
\end{equation}
where $c_k(q)$ is the $q$th element of $ \widehat{\mathbf c}(f_k)$. The maximum values of $ang(l_{\beta_{k}})$ and $vel(l_{v_{k}})$ will occur when $ l_{\beta_{k}} = \frac{M_{R}}{\lambda} d \sin \beta_{k} $ and $l_{v_{k}} = f_{k} Q T_{s} $ , respectively. Therefore, the DoA and the velocity estimations of the $ k $-th target are obtained by
\begin{equation}
	\left\{
	\begin{aligned}
		& \widehat{\beta}_{k} = \arcsin (\dfrac{\lambda L_{\beta_{k}}}{M_{R}d}),\\
		& \widehat{v}_{k} = \dfrac{\lambda L_{v_{k}}}{Q T_{s}},
	\end{aligned}
	\right.
\end{equation}
where $ L_{\beta_{k}} = \arg \max ang(l_{\beta_{k}}) $ and $ L_{v_{k}} = \arg \max vel(l_{v_{k}}) $. Since different subcarrier sets are used to detect different targets, we use the corresponding subcarrier data in $ \widehat{\mathbf b}(\tau_k) $ to estimate $ r_{k} $
\begin{equation}
	\begin{aligned}	
		ran(l_{r_k}) & = {\rm IDFT} [ \widehat{\mathbf b}(\tau_k)]
		 \\ & = \frac{1}{|\mathcal N_k|} \sum _{n\in \mathcal N_k}b_{k}(n) \exp \left(j\frac{2 \pi}{|\mathcal N_k|}(n-1)l_{r_k} \right),
	\end{aligned}
\end{equation}
where $b_k(n)$ is the $n$th element of $ \widehat{\mathbf b}(\tau_k)$, and $|\mathcal N_k|$ is the number of the subcarrier allocated to the $k$-th target. Accordingly, the range estimation can be obtained by the following formula
\begin{equation}
	\begin{aligned}	
		\widehat{r}_{k} = \dfrac{cL_{r_k}}{N_k \Delta f},
	\end{aligned}
\end{equation}
where $ L_{r_k} = \arg \max ran(l_{r_k}) $.

\section{Simulation Result}
In this section, we present some simulation results about the performance with different parameter configurations in a bastatic integrated MIMO-OFDM system. We set $ d = \lambda /2 $, $f_c=3$GHz, $\Delta f=15$kHz, $N=128$, $Q=32$ and $T_{cp}=4.7\mu s$. The SNR is measured by
\begin{equation}
	\begin{aligned}	
		{\rm SNR} _{ \left[ \rm dB \right] } = 10 \log \dfrac{\Vert \mathbf{S}_{1} \Vert_{F}^{2}}{ \mathbb E \{ \Vert \mathbf{V}_{1} \Vert_{F}^{2} \} } .
	\end{aligned}
\end{equation}
The path-loss of communication receiver $ A_k^{'}$ is
$ \sqrt{\frac{\lambda^2}{(4\pi)^2(r_k^1)^{2.5}}}$
and the the attenuation coefficient $ \widetilde{A}_{k} $ is
$
\sqrt{\frac{\lambda^2RCS}{(4\pi)^3(r_k^1)^{2}(r_k^2)^{2}}},
$
where $ RCS = 0.1 $.
\begin{figure}
	\centerline{\includegraphics[scale=0.45]{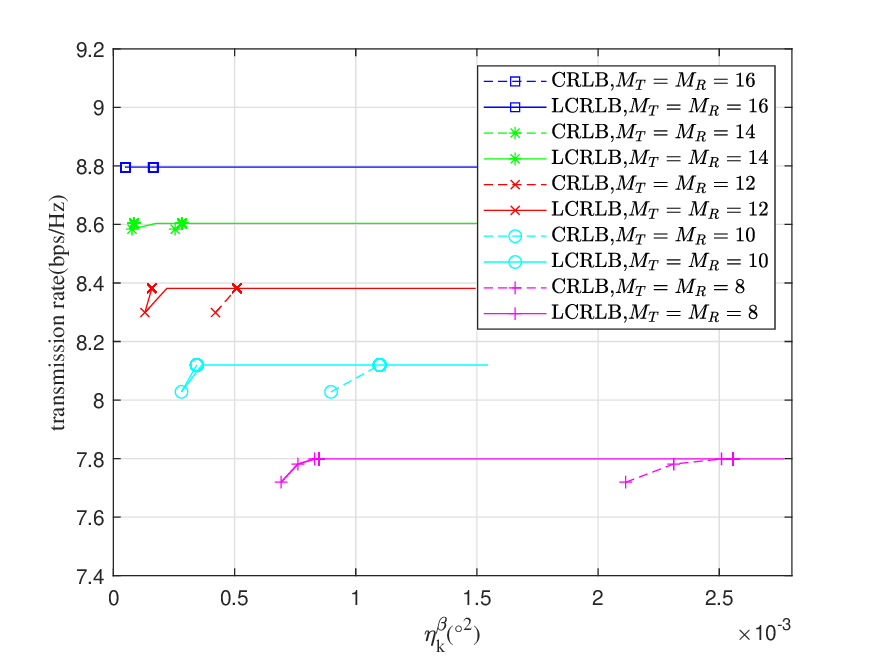}}
	\vspace{-1.3pt}
	\caption{Tradeoff performance between transmission rate and CRLB constraint of DoA with different $M_{T}$ and $M_R$.}
	\label{fig3}
	\vspace{-12pt}
\end{figure}
\begin{figure}
	\centerline{\includegraphics[scale=0.45]{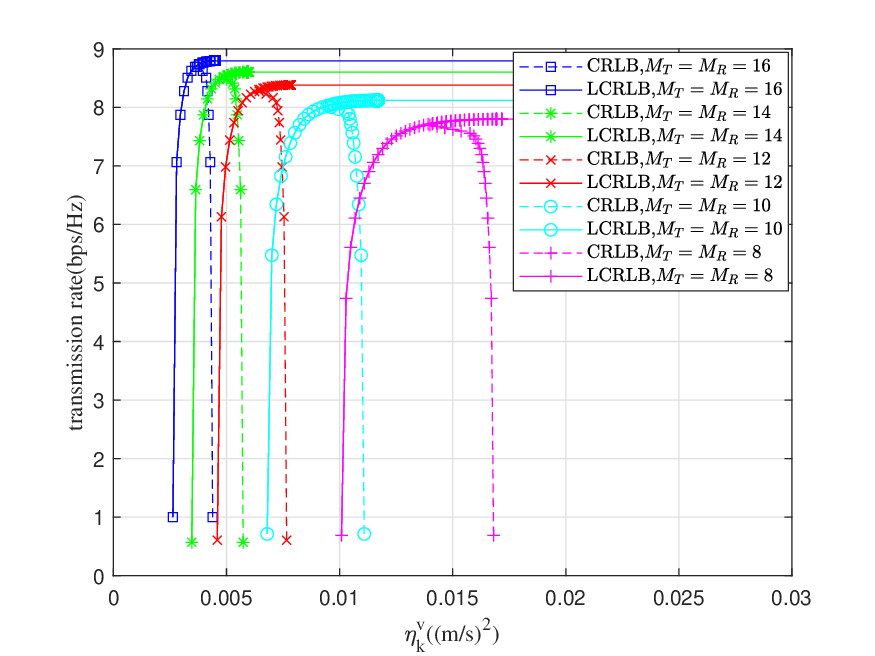}}
	\vspace{-1.5pt}
	\caption{Tradeoff performance between transmission rate and CRLB constraint of velocity with different $M_T$ and $M_R$.}
	\label{fig4}
	\vspace{-12pt}
\end{figure}
\begin{figure}
	\vspace{-15pt}
	\centerline{\includegraphics[scale=0.45]{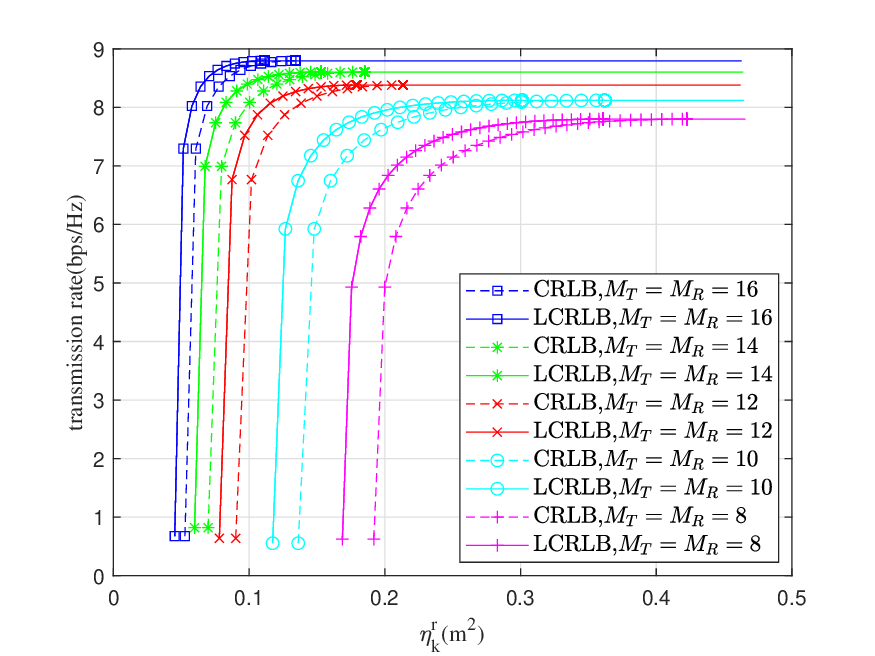}}
	\vspace{-2pt}
	\caption{Tradeoff performance between transmission rate and CRLB constraint of distance with different $M_T$ and $M_R$.}
	\label{fig5}
\end{figure}

\subsection{Tradeoff between Communication and Radar Performances}\label{5.1}

%

Power allocation has a significant influence on both communication rate and radar sensing. Fig.\ref{fig3}-\ref{fig5} illustrates the achieved tradeoff between the spectral efficiency and CRLBs by optimizing the transmit power allocation among different subcarriers. In order to observe the different effects of angle, velocity and distance estimation performance constraints on the data transmission rate, we fix the CRLB constraints of other two parameters in Fig.\ref{fig3}-\ref{fig5}. In the case of multi-objective, the CRLB constraints of the different targets, such as $ \{\eta_k^v\}_{k=1}^K $ are set to be the same. We set $K=2$ and the position and velocity parameters of the two targets are $ \{\alpha_{k} \}_{k = 1}^{2} = \{ 30^{\circ}, \ 5^{\circ} \}, \{\beta_{k} \}_{k = 1}^{2} = \{ 10.23^{\circ}, \ 30.34^{\circ} \}, \{r_{k} \}_{k = 1}^{2} = \{ 1000.42,\ 1050.75 \} {\rm m}, \{v_{k} \}_{k = 1}^{2} = \{ 19.21,\ 20.36 \} {\rm m/s} $. Subcarriers are divided into two groups in advance for two targets. The total transmit power of the transmitter is $ p_T = 5 \rm W $ and the PSD of communication receiver and radar receiver are $ n'_0 = 1e{-18} $ and $ n_0 = 1.55e{-22} $. In Fig.\ref{fig3}, the solid lines represent the relationship between the constraint $ \eta_k^\beta $ and the communication rate with different numbers of antennas. With the optimized power allocation, we can calculate CRLB ($-*-$) and LCRLB (\:\sout{$ \;\;\; * \;\;\; $}\:) of DoA estimation with \eqref{eq32}-\eqref{eq34}. As shown in Fig.\ref{fig3}, when $M_T=8, M_R=8$, the minimum achievable LCRLB of DoA is $ 6.92e{-4} $. When $ \eta_k^\beta < 6.92e{-4} $, the problem in $ {\eqref{optimize-problem}}$ is infeasible. It also shows that increasing the number of antennas helps to reduce CRLB of DoA and the gap between CRLB and LCRLB. This is because the CRLB matrix in (29) will gradually approach the block diagonal matrix with the increase of the number of antennas. We can also find that increasing $\eta_k^\beta$ has little effect on the CRLB and transmission rate, i.e., even if the constraint ({\ref{optimize-problem}c}) is removed from the optimization problem, the desired accuracy of DoA estimation can still be obtained. It can be observed from Fig.\ref{fig3}, the CRLB constraints of velocity and distance estimation have a greater impact on the transmission rate compared with the DoA estimation constraint.

From Fig.\ref{fig4},\ref{fig5}, we can see that the communication rate continues to increase with the increase of $\eta_k^v $ or $\eta_k^r$, and finally reaches a constant value. For example, when $M_T = M_R=8$, the minimum LCRLB in Fig.\ref{fig4} is $ 0.01 \rm $. As $\eta_k^v$ increases, the transmission rate and the LCRLB also increase gradually. When $ \eta_k^v = 0.017 $, the LCRLB reaches its maximum value, i.e., the LCRLB no longer increases with the increasing of $\eta_k^v$. That is, the LCRLB of velocity estimation can still be about 0.017 without the constraint $ \widetilde{\mathbf C}_k^v \succeq 0$ in $({\ref{optimize-problem}b})$. With a slight loss of velocity estimation performance, a significant improvement of communication rate can be obtained. Similarly, we can obtain an obvious increase in communication rate by reducing the LCRLB of distance by 0.2, as shown in Fig.\ref{fig5}. Note that the CRLB of velocity ($-+-$) in Fig.\ref{fig4} does not increase monotonically with the increase of $\eta_{k}^{v}$ because when $\eta_{k}^{v} > 0.01$, the gap between CRLB of velocity and LCRLB of velocity decreases gradually. Fig.\ref{fig3}-\ref{fig5} demonstrates that as the number of transmit antennas increases, the communication rate increases, the CRLBs of DoA, velocity and distance estimation decreases and the gap between the LCRLB and the CRLB becomes smaller.

\subsection{The Effect of Power Allocation on Estimation Performance}\label{5.2}

%

\begin{figure}
	\centerline{\includegraphics[scale=0.45]{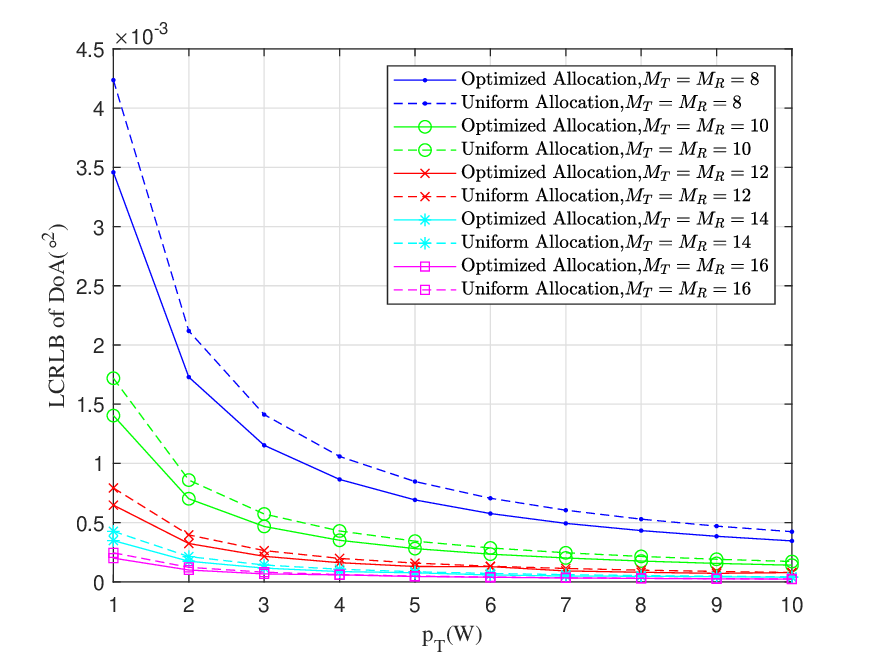}}
	\vspace{-1.5pt}
	\caption{The LCRLB of DoA vs. power constraint $p_T$ with different power allocation schemes.}
	\label{fig6}
	\vspace{-10pt}
\end{figure}
\begin{figure}
	\centerline{\includegraphics[scale=0.45]{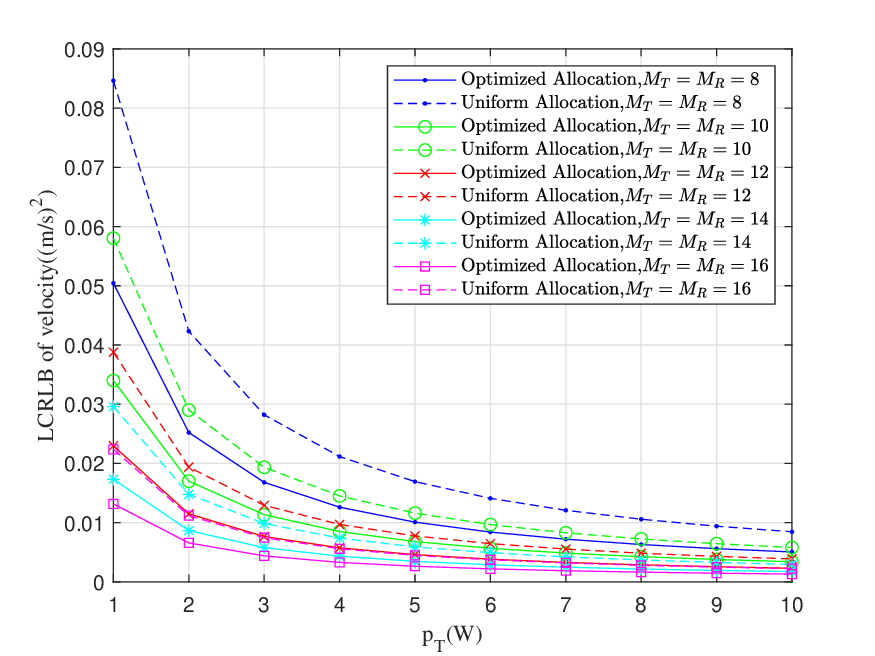}}
	\vspace{-2.4pt}
	\caption{The LCRLB of velocity vs. power constraint $p_T$ with different power allocation schemes.}
	\label{fig7}
	\vspace{-17pt}
\end{figure}
\begin{figure}[t]
	\centerline{\includegraphics[scale=0.45]{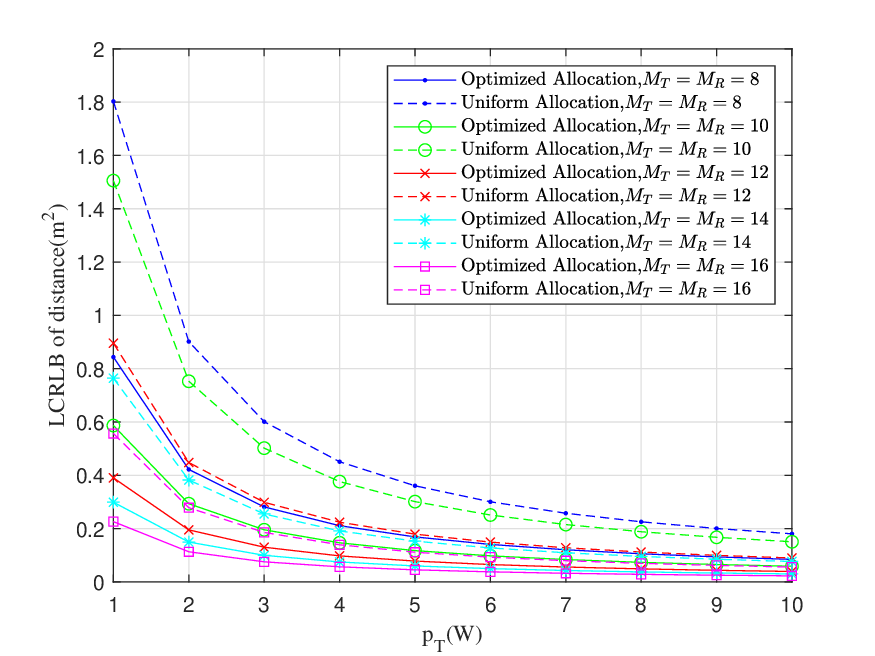}}
	\vspace{-5pt}
	\caption{The LCRLB of distance vs. power constraint $p_T$ with different power allocation schemes.}
	\label{fig8}
	\vspace{-10pt}
\end{figure}
To observe the effect of transmit power allocation on the estimation performance, we retain the minimum achievable LCRLBs when the problem in $ {\eqref{optimize-problem}}$ is feasible. Fig.\ref{fig6}-\ref{fig8} shows the minimum achievable LCRLBs with the optimized power allocation and the uniform power allocation schemes for different $ p_T $.  Except for $p_T$, other parameters of system are the same as those in \ref{5.1}. In Fig.\ref{fig6}, the solid lines represents the LCRLB of DoA after optimizing the power allocation and the dashed lines denote the LCRLB of DoA with the uniform power allocation. For the same number of antennas, we can see that the solid line is lower than the dashed one and as the total power $ p_T $ increases, the gap between them shrinks. This shows that optimized power allocation can better improve the accuracy of angle estimation at low SNR. In the high SNR region, LCRLB of DoA with uniform power allocation has been very small, so that the improvement resulted from optimized power allocation is not significant. As the total power increases, the decrease in the LCRLB also tends to be flattened. In Fig.\ref{fig6},\ref{fig7}, the effect of the power allocation on the LCRLBs of velocity and distance is similar to that in Fig.\ref{fig6}. Compared with results in Fig.\ref{fig6}, the gap between the solid and dashed lines is greater. It illustrated that the detection performance of velocity and distance is significantly improved after optimizing the transmit power of the subcarriers. Therefore, in order to improve the detection performance, it is necessary to optimize the power allocation.

\begin{figure}
	\centerline{\includegraphics[scale = 0.45]{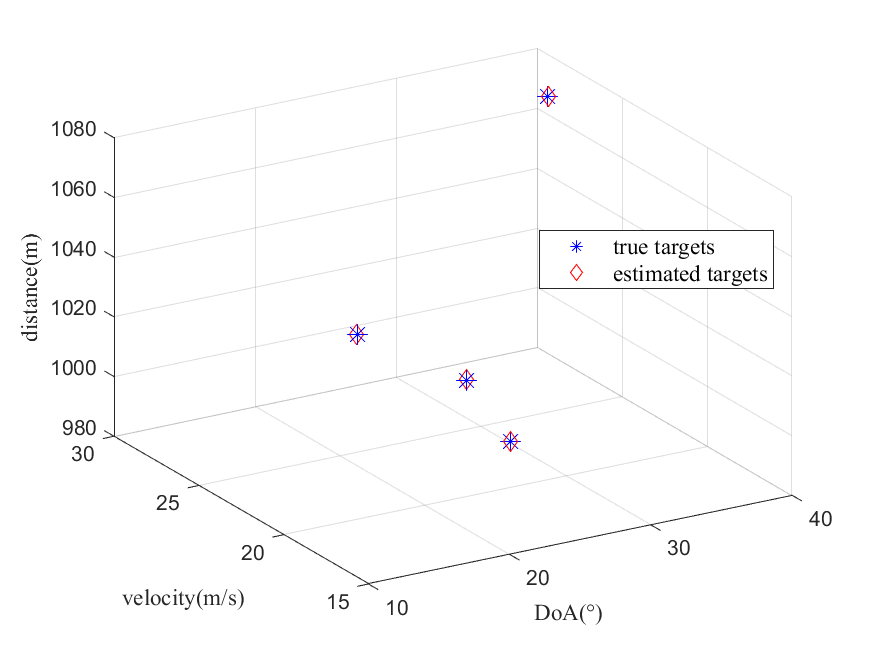}}
	\vspace{-2pt}
	\caption{3D parameters estimation results.}
	\label{fig9}
	\vspace{-10pt}
\end{figure}
\begin{figure}
	\centering
	\subfigure[single-target]
	{
		\begin{minipage}{0.46\linewidth}
			\centerline{\includegraphics[scale=0.32]{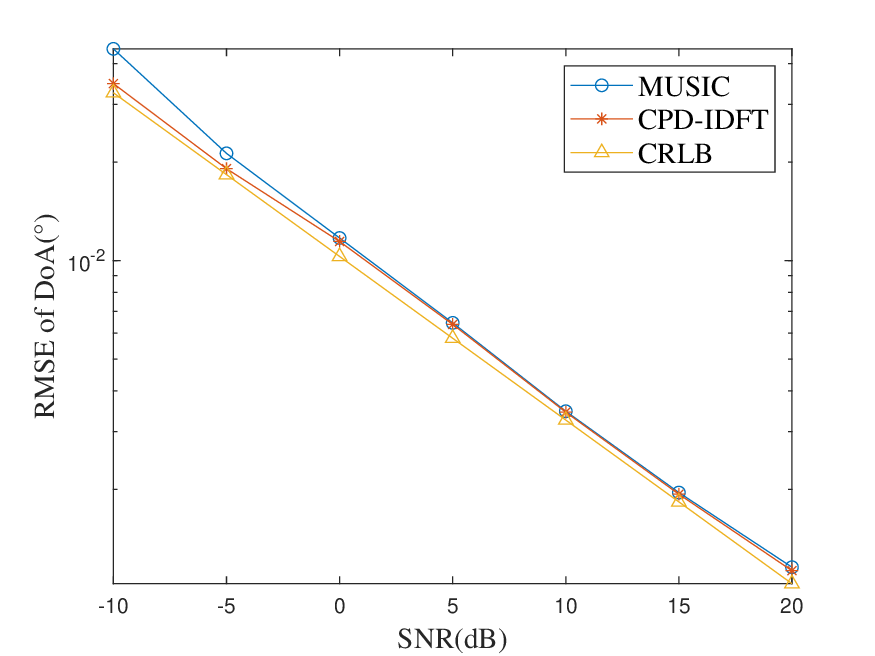}}
			\centerline{\includegraphics[scale=0.32]{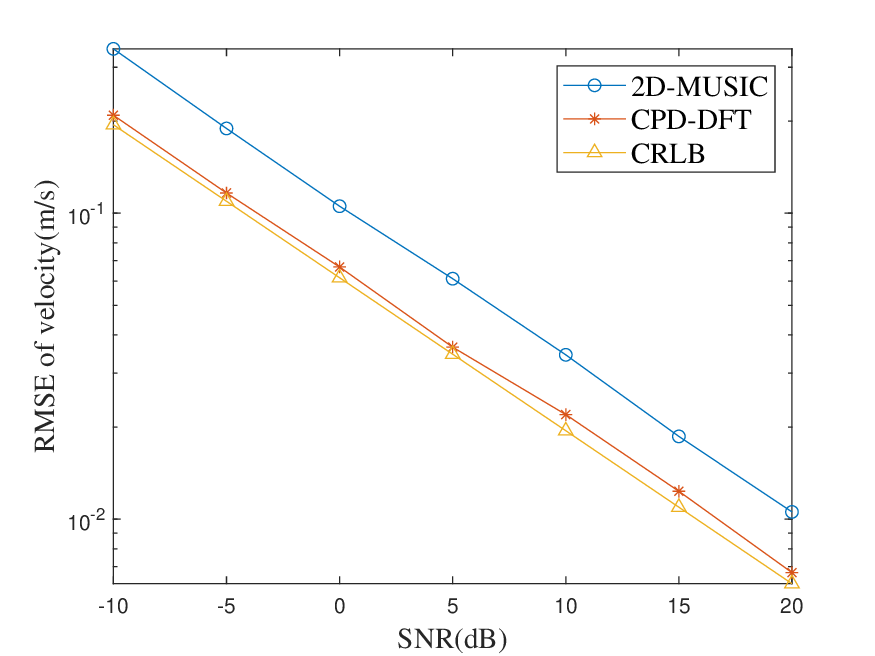}}
			\centerline{\includegraphics[scale=0.32]{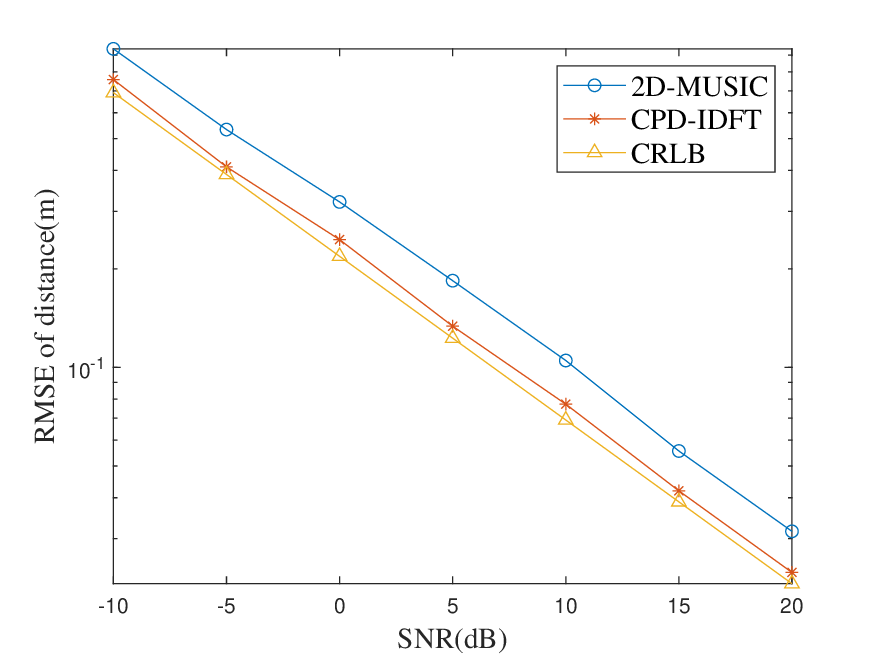}}
		\end{minipage}
	}
	\subfigure[multi-targets]
	{
		\begin{minipage}{0.46\linewidth}
			\includegraphics[scale=0.32]{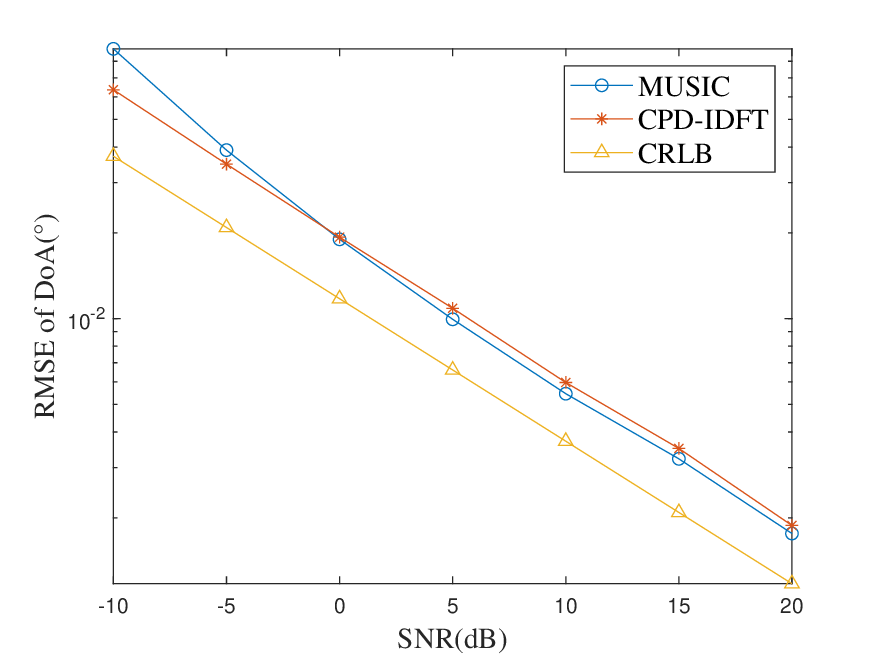}
			\includegraphics[scale=0.32]{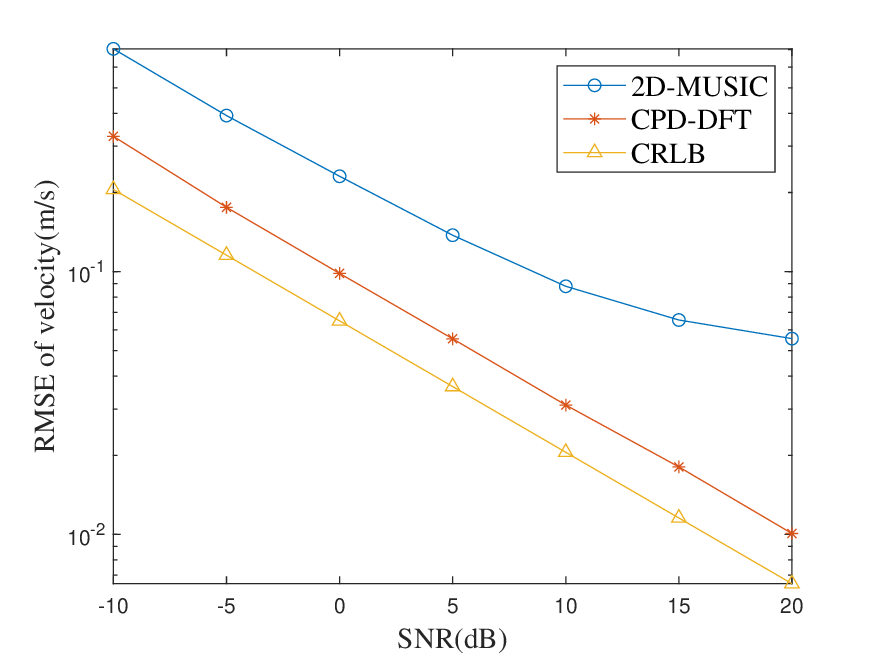}
			\includegraphics[scale=0.32]{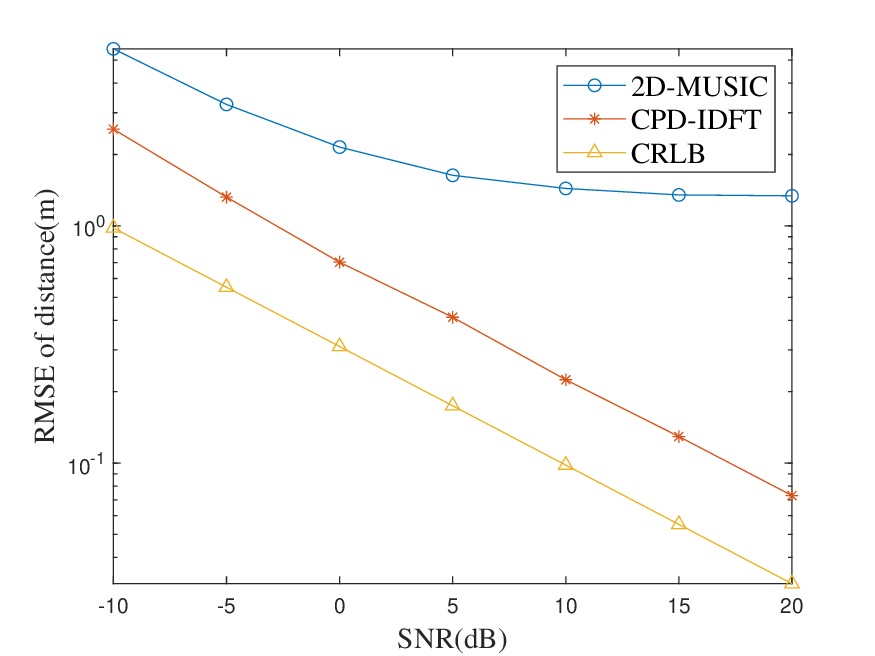}
		\end{minipage}
	}
	\caption{Estimation performance of different estimation methods.}
	\label{fig10}
	\vspace{-15pt}
\end{figure}

\subsection{Estimation Performance Based on CPD}
In the integrated communication-radar system we study, the signal model can be associated with the 3D-tensor. We decouple the target parameters into three factor matrices by CPD, and utilize the phase characteristics to estimate the DoA, velocity and distance of the targets in parallel. In this subsection, we demonstrate the estimation performance of the CPD-based algorithm by comparing the proposed algorithm and the traditional algorithm. Meanwhile, CRLBs are made as a benchmark to show the difference between the CPD-based estimation method and the theoretical minimum estimation error. In the following two cases, we demonstrate the estimation performance with uniform power allocation and set $ M_T = M_R = 16 $, $ p_T = 5 \rm W $.

Case 1: Fig.\ref{fig9} demonstrates that the position and velocity of multi-targets can be accurately estimated by our proposed CPD-based method. Suppose that subcarriers are evenly allocated to four targets and the DoDs and DoAs of them are $ \{\alpha_{k} \}_{k = 1}^{4} = \{ 5^{\circ}, \ 18^{\circ}, \ -26^{\circ}, \ -22^{\circ} \} $ and $ \{\beta_{k} \}_{k = 1}^{4} = $ $ \{ 10.23^{\circ}, \ 30.09^{\circ}, \ 22.56^{\circ}, \ 38.85^{\circ} \} $ respectively. The velocity and the range of targats are $ \{v_{k} \}_{k = 1}^{4} = \{ 15.86,\ 23.34,\ 19.67,\ 28.44 \} {\rm m/s} $ and $ \{r_{k} \}_{k = 1}^{4} = \{ 1060.35,\ 980.24,\ 1020.46,\ 1070.52 \} {\rm m} $. In Fig.\ref{fig9}, we plot the accurate locations and velocities (marked with `$ \ast $') and estimated locations and velocities (marked with `$ \diamond $') of four targets. We can see that accurate values and estimated values can  match well in three dimensions.

%

Case 2: To demonstrate the estimation performance of the proposed CPD-based method, we consider single-target and multi-targets scenarios in Fig.\ref{fig10}.
In the single-target scenario of Fig.\ref{fig10}(a), the DoD, DoA, velocity and range of the target are $ \{\alpha_{k} \}_{k = 1}^{1} = \{ 5^{\circ} \} $,
$ \{\beta_{k} \}_{k = 1}^{1} = \{ 10.23^{\circ} \} $, $ \{v_{k} \}_{k = 1}^{1} = \{ 19.21 \} {\rm m/s} $ and $ \{r_{k} \}_{k = 1}^{1} = \{ 960.42 \} {\rm m} $. In multi-targets scenario of Fig.\ref{fig10}(b), the DoD, DoA, velocity and distance of the targets are $ \{\alpha_{k} \}_{k = 1}^{2} = \{ 5^{\circ}, 30^{\circ} \} $, $ \{\beta_{k} \}_{k = 1}^{2} = \{ 10.23^{\circ}, 30.34^{\circ}\} $, $ \{v_{k} \}_{k = 1}^{2} = \{19.21, 25.36 \} {\rm m/s} $ and $ \{r_{k}\}_{k = 1}^{2} = \{960.42, 1120.75\} {\rm m} $. In Fig.\ref{fig10}(a), we compare the RMSE (root mean squared error) of estimated parameters obtained by different parameters estimation methods under different SNR in the single target scenario. In the picture of DoA estimation, we can see that the RMSE of CPD-based method is almost the same with that of MUSIC method when $ \rm {SNR} > 0 \rm{dB} $. However, CPD-based method can obtain lower RMSE which is closer to CRLB in the low SNR region. In the pictures of velocity and distance estimation, the estimation  performance of CPD-based method is much closer to CRLBs than that of 2D-MUSIC method and can get about $5 \rm dB $ and $ 2\rm dB$ improvements of velocity and distance estimation under the same RMSE respectively. In Fig.\ref{fig10}(b), the performance metric corresponding to y coordinate is the average RMSE of two targets. In contrast with the results in single target scenario, in the multi-target scenario, the gap between RMSE of CPD-based method and CRLB increases,  partially because of the mutual interference between echo signals of different
\begin{figure}
	\centerline{\includegraphics[scale=.45]{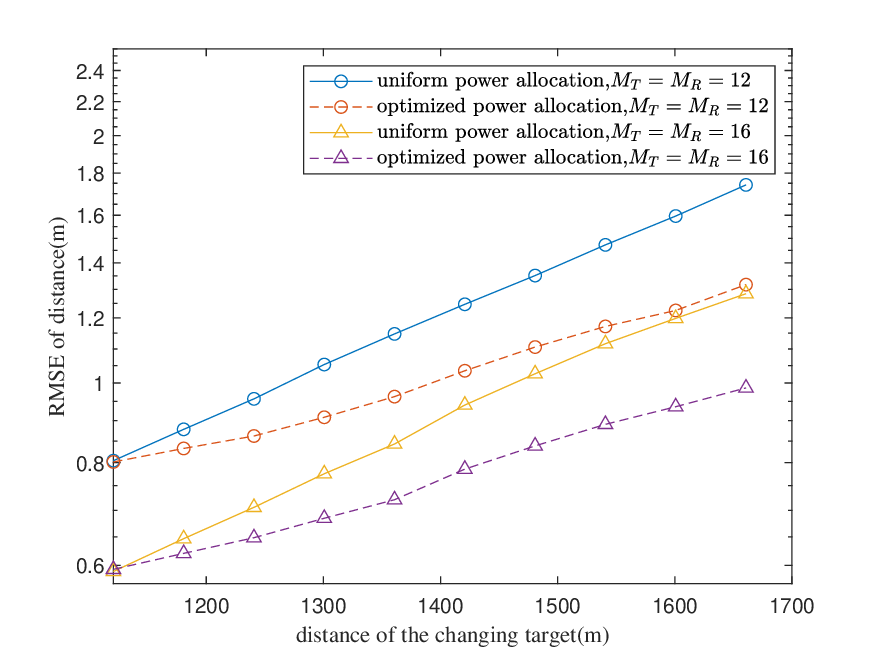}}
	\vspace{-2pt}
	\caption{ The estimation performance with different power allocation schemes.}
	\label{fig11}
	\vspace{-18pt}
\end{figure}
targets. In terms of DoA estimation, the CPD-based method can still achieve a better estimation accuracy than the MUSIC-based method in low SNR region. Moreover, in the multi-target scene, the CPD-based method has greater advantages over 2D-MUSIC in distance and velocity estimation. Specially, in the velocity and distance estimation, RMSE of 2D-MUSIC method levels off at high SNR due to the interference between targets.

Since the distance of the target is the main factor affecting the estimation performance, in the above scenario with $K=2$, we fix the distance of one target $r_1^1 + r_1^2 = 860.42$m, and change the distance of another target $r_2^1+r_2^2$ from $ 1120.75 \rm m $ to $ 1690.75 \rm m $. At each distance, we set $ \eta_k^r $ in ({\ref{optiz-problem}c}) to be the minimum value which makes the problem in $ {\eqref{optimize-problem}}$ feasible and adopt the optimal subcarrier power allocation obtained at the transmitter. The total transmit power and the PSD are the same as those in \ref{5.1}.
At the radar receiver, we adopt the CPD-based method to estimate the distance of the changing target. Fig.\ref{fig11} shows the RMSE of distance estimation based on CPD method under uniform power allocation and optimized power allocation schemes with different numbers of antennas. As the distance difference between targets increases, optimal power allocation achieves higher distance estimation accuracy compared with uniform power allocation.


\section{Conclusion}
In this paper, we investigate an integrated communication-radar MIMO-OFDM system, where the base station communicates with targets, and the echo signals are utilized to estimate the positions of targets at the radar receiver. We first derive the closed-form expression of the lower bound of CRLB of target parameters. Then, considering the power allocation over subcarriers has different effects on communication performance and estimation performance, we optimize the power allocation to achieve the optimal tradeoff between transmission rate and estimation performance. Finally, we propose a parallel estimation scheme based on CPD, which can obtain better estimation accuracy than traditional estimation method. Moreover, utilizing the optimized power allocation, the estimation performance of the CPD-based method can be further improved.

\end{document}